\documentclass[a4paper,11pt]{article}
\pdfoutput=1 

\usepackage{jheppub,bm} 

\usepackage[T1]{fontenc} 
\usepackage[utf8]{inputenc}

\usepackage{slashed} 
\usepackage{lipsum}
\usepackage{mathtools} 
\usepackage{bbm}
\usepackage{float}
\usepackage{subcaption}
\usepackage{xcolor}
\usepackage{adjustbox}
\usepackage{amsmath}
\usepackage{tikzfeynman} 
\usepackage[T1]{fontenc} 
\usepackage[utf8]{inputenc}

\def\eps{\epsilon}
\def\mmu2om2{\left (\frac{\mu^2}{m^2} \right )^{\!\eps}}

\def\spa#1.#2{\langle #1 #2\rangle}
\def\spb#1.#2{[ #1 #2]}
\def\spab#1.#2.#3{\langle #1 |#2| #3] }

\newcommand{\cmark}{{\color{green}\Large{\checkmark}}}
\newcommand{\xmark}{{\color{red}$\boldsymbol{\times}$}} 

\title{Electroweak corrections to $gg\rightarrow \gamma\gamma$} 

\author[a]{Gabriele Fiore,}
\affiliation[a]{
	Institute for Theoretical Physics, ETH, 8093 Zürich, Switzerland}
\author[b]{Ciaran Williams}    

\affiliation[b]{Department of Physics,\\University at Buffalo, The State University of New York, Buffalo
14260 USA}

\emailAdd{gfiore@ethz.ch}
\emailAdd{ciaranwi@buffalo.edu}

\begin{abstract}{
	We present the electroweak corrections for the production of a photon pair through gluon fusion, focusing on the contribution from the first two generations of quarks. 
	The two-loop amplitude is calculated using a series of projection operators which define scalar form factors. In order to evaluate the Master Integrals which appear in this process we employ both generalized polylogarithms and Chen-iterated integrals. In order to perform a phenomenological study we develop a semi-numerical evaluation of the Master Integrals employing a fitting procedure to speed up the evaluation of burdensome higher weight contributions. We present results for the LHC, finding corrections of around a couple of percent to the leading order $gg \rightarrow \gamma\gamma$ process. Our results are implemented into the parton-level Monte Carlo code \texttt{MCFM}. }
\end{abstract}

\begin{document} 
\maketitle
\flushbottom


\section{Introduction}
\label{section:intro}

The production of photon pairs (diphotons) at hadron colliders has been studied experimentally for decades~\cite{Bonvin:1988yu,Alitti:1992hn,Abe:1992cy,Abazov:2010ah,Aaltonen:2012jd,Chatrchyan:2011qt,Aad:2012tba,Aad:2013zba,Aaltonen:2011vk,Chatrchyan:2013mwa,Chatrchyan:2014fsa,Marini:2016zjr,CMS:2018qao,ALICE:2019rtd,ATLAS:2019buk,ATLAS:2019iaa,CMS:2019jlq,ATLAS:2021mbt,ATLAS:2023yrt}, establishing the process as a standard candle for particle physics at colliders. The rich data set has also encouraged extensive study of the process theoretically, ranging from early studies at Next-to-Leading Order (NLO) in QCD~\cite{Aurenche:1987fs,Baer:1990ra,Berger:1990et,Gordon:1994ut,Balazs:1997hv,Vogelsang:1995bg,Binoth:1999qq}, Next-to-Next-to Leading Order (NNLO) predictions~\cite{Bern:1994fz,Anastasiou:2002zn,Campbell:2016yrh,Catani:2011qz} to $q_T$-resummed results at N$^3$LL~\cite{Neumann:2021zkb}. Significant progress on the complete N3LO calculation has also been made, with NNLO predictions for diphoton plus jet available~\cite{DelDuca:2003uz,Gehrmann:2013bga,Badger:2013ava,Chawdhry:2021hkp} and three-loop amplitudes for $q\bar{q} \to \gamma\gamma$~\cite{Caola:2020dfu}. Further recent progress for QCD corrections have been undertaken, with recent results at two-loop focusing on heavy-quark contributions~{\cite{Coro:2025vgn,Ahmed:2025osb}.
	
	The dominant production mechanism for the hadronic production of diphotons proceeds through quark anti-quark fusion, i.e. $q\bar{q} \to \gamma\gamma$, with the resulting cross section starting at $\mathcal{O}(\alpha^2)$. At NNLO in QCD a new partonic channel $gg\rightarrow \gamma\gamma$ opens up, which, due to the large flux at LHC energies contributes a significant fraction of the total cross section, despite beginning at $\mathcal{O}(\alpha^2\alpha_s^2)$. Given its significant contribution to the cross section, the calculation has attracted a fair amount of theoretical attention over the years. After a pioneering study in the late 1980's at LO~\cite{Glover:1988fe}, NLO (QCD) corrections were completed in the earlier 2000's~\cite{Bern:2001df,Bern:2002jx}. These have been extended to include top-quark effects more recently~\cite{Chen:2019fla,Maltoni:2018zvp}. Three-loop QCD amplitudes for this process have been calculated~\cite{Bargiela:2021wuy} and master integrals for the EW corrections have been calculated in ref.~\cite{Fiore:2023myh}.

	Another reason why the $gg\rightarrow\gamma\gamma$ process is of interest is its role in Higgs physics. At $\mathcal{O}(\alpha^3\alpha_s^2)$ a Higgs boson can be created through the fusion of two gluons (through a heavy-quark loop) with a subsequent decay of the Higgs to photon pairs. Given the excellent resolution of the LHC experiments in measuring the invariant mass of photon pairs, this channel was one of the two in which the Higgs boson was initially discovered~\cite{Chatrchyan:2012ufa,Aad:2012tfa}.

	However, the significance of the diphoton channel extends well beyond its historical role in the discovery. It remains a central probe for precision studies of the Higgs boson. After the discovery phase, research focus transitioned to the in-depth exploration of the Higgs boson properties and its interaction with the other components of the Standard Model (SM). A major thrust of the LHC physics program has been dedicated to measuring key observables such as the Higgs mass, its total decay width $(\Gamma_H)$, and the couplings to other SM particles. The diphoton channel, has enabled a precise determination of the Higgs boson mass with an uncertainty of approximately 140 MeV~\cite{ATLAS:2023owm}. Complementary to this, the $ZZ^*$ decay channel achieves an uncertainty around 180 MeV~\cite{ATLAS:2022zjg} in the fully leptonic decay channel. When these two channels are combined, the overall uncertainty on the Higgs boson mass measurement is reduced to about 110 MeV~\cite{ATLAS:2023oaq}.
	
	One area where Higgs to diphotons is particularly interesting is in regards to direct constraints on $\Gamma_H$ at the LHC, which proceeds by exploiting the interference between the resonant Higgs-mediated amplitude and the continuum background in the $gg\rightarrow\gamma\gamma$ process. The interference induces a distortion in the diphoton invariant mass distribution, manifested as a shift in the peak position. Critically, the magnitude of this shift is directly sensitive to the total decay width of the Higgs boson.
	Early studies at leading order (LO) in QCD demonstrated that the interference effect could induce a mass shift in the range of 100–200 MeV under typical LHC conditions~\cite{Martin:2012xc}. These initial analyses were subsequently extended to incorporate contributions from subdominant processes, such as $q(\bar{q})g$ and $q\bar{q}$ initial states, introducing an additional shift of around 30 MeV with an opposite sign to that of the gluon-fusion channel~\cite{deFlorian:2013psa}. Building upon these foundations, the impact of NLO QCD corrections was later computed~\cite{Dixon:2013haa}, revealing a significant modification of the mass shift, with corrections exceeding 40\%. 
		
	Further refinements have been explored in the literature~\cite{Dixon:2013haa,deFlorian:2013psa,Cieri:2017kpq,Campbell:2017rke}, including the incorporation of small transverse momentum ($p_T$) resummation~\cite{Cieri:2017kpq} and the role of additional hard QCD radiation~\cite{Coradeschi:2015tna}. Notably, these studies highlighted the dominant role of the low-$p_T$ region of the diphoton system in driving the interference-induced mass shift~\cite{Dixon:2013haa,Coradeschi:2015tna,Martin:2013ula}. For several years, the main bottleneck for further theoretical progress in this direction has been the absence of the necessary multi-loop amplitudes for the continuum $gg\to\gamma\gamma$ process. However, this has recently been overcome, thanks to substantial advances in the computation of QCD scattering amplitudes. In particular, the availability of three-loop helicity amplitudes for $gg\to\gamma\gamma$~\cite{Bargiela:2021wuy}, two-loop amplitudes for $\gamma\gamma+j$ production~\cite{Agarwal:2021vdh,Badger:2021imn}, and the inclusion of finite top-quark mass effects at two loops~\cite{Becchetti:2025rrz,Davies:2025out}, now collectively open the possibility of achieving a complete NNLO assessment of the signal–background interference and the associated mass shift. A significant recent advance, representing the current state of the art for NNLO corrections, is the evaluation carried out in the soft-virtual approximation~\cite{Bargiela:2022dla}. The authors report an additional $\approx$1.7\% reduction in the cross section and a further $\approx$30\% decrease in the mass-shift effects, highlighting that these analyses remain sensitive to higher-order corrections.

	Thanks to these developments, and with the steadily improving precision of both experimental measurements and theoretical predictions, the incorporation of electroweak corrections to the interference process becomes increasingly pressing.  In this paper, we make the first step towards this goal by computing the $\mathcal{O}(\alpha)$ electroweak corrections to the continuum $gg\to\gamma\gamma$ process, focusing specifically on the effects from the first two generations of quarks.
	
	This paper proceeds as follows, an outline of our calculation is presented in Section~\ref{sec:overview}. Section~\ref{sec:MIeval} provides a detailed description on the checks and validations we performed on our calculation, as well as a description of how we evaluated the Master Integrals numerically. Our results are presented in Section~\ref{sec:Results}. 


\section{Overview}
\label{sec:overview}

This section outlines the calculation of the electroweak corrections to $gg\rightarrow \gamma\gamma$ presented in this paper. The calculation broadly consists of two stages, amplitude calculation and renormalization, followed by  Master Integral (MI) evaluation. From a phenomenological point of view the last point becomes the most intricate when evaluating the process inside of a Monte Carlo computer code, and we therefore discuss this aspect of the calculation in a dedicated section (Section~\ref{sec:MIeval}). This section will present an initial overview of the calculation of the amplitude and subsequent UV renormalization. 

\subsection{Process overview}

The lowest order $gg \rightarrow \gamma\gamma$ amplitude proceeds through a closed loop of quarks as shown in Fig.~\ref{fig:LO}. This amplitude self-interferes and, therefore, first enters the perturbative expansion at NNLO in QCD (i.e. as an $\mathcal{O}(\alpha_s^2)$ correction). 
\begin{figure}[h]
	\begin{center}
		\begin{tikzpicture}[line width=1 pt, scale=1]
			\draw[gluon] (-2.3,1) -- (-1,1);
			\draw[gluon] (-2.3,-1) -- (-1,-1);
			
			\draw[fermion2] (-1,1) -- (1,1);
			\draw[fermion2] (1,-1) -- (-1,-1);
			\draw[fermion2] (-1,-1) -- (-1,1);
			\draw[fermion2] (1,1) -- (1,-1);
			
			\draw[vector] (1,1) -- (2.3,1);
			\draw[vector] (1,-1) -- (2.3,-1);	
			
			\node at (-2.5,1) {$g$};
			\node at (-2.5,-1) {$g$};			
			\node at (2.5,1) {$\gamma$};
			\node at (2.5,-1) {$\gamma$};

		\end{tikzpicture}%
	\end{center}
	\caption{Leading Order diagram for the
		\label{fig:LO}
		$gg\to\gamma\gamma$ process.}
\end{figure}
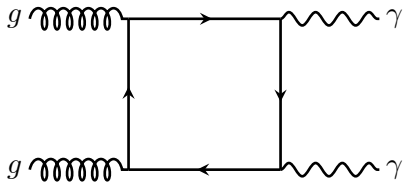
The kinematics of the process can be defined as follows: 
\begin{equation}\label{eq:process}
	0 \rightarrow g(p_1) + g(p_2)+ \gamma(p_3) + \gamma(p_4),
\end{equation}
where we consider all external momenta to be outgoing, and light-like ($p_i^2 =0$).  It is convenient to parameterize the scattering in terms of the following Mandelstam invariants
\begin{align}\label{eq:mandelstams}
	s\! =\! (p_1\! +\! p_2)^2 = 2 p_1 \cdot p_2, \nonumber\\\; 
	t  \! =\! (p_2 \!+ \!p_3)^2 =2 p_2 \cdot p_3,\nonumber\\\;
	u \!=\! (p_1\! +\! p_3)^2 = 2 p_1 \cdot p_3.
\end{align}
Following the discussion in ref.~\cite{Caola:2020dfu,Bargiela:2021wuy}, we introduce a set of projection operators which allows us to decompose the amplitude into scalar form factors multiplying unique tensor structures. We begin by writing the scattering amplitude for the process in Eq.~\ref{eq:process} as:
\begin{equation}
	\label{amplitude0}
	\mathcal{A}_{gg\to\gamma\gamma} = \varepsilon_{\mu}\varepsilon_{\nu}\varepsilon_{\rho}\varepsilon_{\sigma}\mathcal{A}_{gg\to\gamma\gamma}^{\mu\nu\rho\sigma},
\end{equation}
where we have extracted the external polarization vector ($\varepsilon$) for each final state boson. The amplitude tensor $\mathcal{A}_{gg\to\gamma\gamma}^{\mu\nu\rho\sigma}$ can be further expanded in-terms of a set of independent tensor structures:
\begin{equation}\label{eq:tensor_decomp}
	\mathcal{A}_{gg\to\gamma\gamma}^{\mu\nu\rho\sigma} = \sum_{i=1}^{N_T}\mathcal{F}_i \: T_i^{\mu\nu\rho\sigma},
\end{equation}
where $N_T$ is the number of elements in the tensor basis and $\mathcal{F}_i$ are the associated form factors. The tensors $T_i$ are objects defined from combinations of the three independent external momenta, $\gamma$-matrices and the metric tensor $g^{\mu\nu}$. To determine the number of independent tensors, and hence the size $N_T$ of the tensor basis, we implement the symmetries of the problem to remove all redundant tensor structures. A summary of these symmetries and properties are as follows:
\begin{equation}
	\begin{aligned}
		p_i\cdot\varepsilon_i &= 0,  && \quad i = 1,2   &&\hspace{1cm} \makebox[5cm][l]{Gluon Transversality,} \\
		p_i\cdot\varepsilon_i &= 0,  && \quad i = 3,4   &&\hspace{1cm} \makebox[5cm][l]{Photon Transversality,} \\
		\varepsilon_i\cdot p_{i+1} &= 0, && \quad i = 1,\dots,4 \quad \text{with } p_5=p_1   &&\hspace{1cm} \makebox[5cm][l]{Cyclic Gauge.}
	\end{aligned}
\end{equation}
This procedure reduces the total number of tensor structures to 10. However given that each massless boson has $2$ polarization states, and an overall parity symmetry, the total number of independent helicity configuration should be 8. Therefore, there exists a redundancy which can further constrain the tensors. By combining the tensor structures that purely depend on the metric tensor, two further tensors can be shown to be linearly dependent on the remaining eight. In summary, we conclude that $N_T=8$ and we can rewrite Eq.~\ref{eq:tensor_decomp} as:
\begin{equation}\label{eq:tensor_decomp2}
	\mathcal{A}_{gg\to\gamma\gamma}^{\mu\nu\rho\sigma} = \sum_{i=1}^{8}\mathcal{F}_i \: T_i^{\mu\nu\rho\sigma}.
\end{equation}
Further, we define the contraction of the tensors with the external polarizations as follows:
\begin{align} \label{eq:Tensors}
	&T_1 = \epsilon_1 \! \cdot \! p_3\; \epsilon_2 \! \cdot \! p_1\; \epsilon_3 \! \cdot \! p_1\; \epsilon_4 \! \cdot \! p_2 \;, \nonumber \\
	&T_2 = \epsilon_1 \! \cdot \! p_3\; \epsilon_2 \! \cdot \! p_1\; \epsilon_3 \! \cdot \! \epsilon_4 , \quad
	T_3 = \epsilon_1 \! \cdot \! p_3\; \epsilon_3 \! \cdot \! p_1\; \epsilon_2 \! \cdot \! \epsilon_4 , \nonumber \\
	&T_4 = \epsilon_1 \! \cdot \! p_3\; \epsilon_4 \! \cdot \! p_2\; \epsilon_2 \! \cdot \! \epsilon_3, \quad T_5 = \epsilon_2 \! \cdot \! p_1\; \epsilon_3 \! \cdot \! p_1\; \epsilon_1 \! \cdot \! \epsilon_4 , \nonumber \\
	&T_6 = \epsilon_2 \! \cdot \! p_1\; \epsilon_4 \! \cdot \! p_2\; \epsilon_1 \! \cdot \! \epsilon_3 , \quad
	T_7 = \epsilon_3 \! \cdot \! p_1\; \epsilon_4 \! \cdot \! p_2\; \epsilon_1 \! \cdot \! \epsilon_2 , \nonumber \\
	&T_8 = \epsilon_1 \! \cdot \! \epsilon_2\;  \epsilon_3 \! \cdot \! \epsilon_4+ \epsilon_1 \! \cdot \! \epsilon_4\;  \epsilon_2 \! \cdot \! \epsilon_3 + \epsilon_1 \! \cdot \! \epsilon_3\;  \epsilon_2 \! \cdot \! \epsilon_4 \;.
\end{align}
We use projector operators in order to isolate individual form factors entering into the expansion in Eq.~\ref{eq:tensor_decomp2}:
\begin{equation}
	\mathcal{F}_{j}=\sum_{pol}\mathcal{P}_{j}\mathcal{A}_{gg\rightarrow\gamma\gamma},
\end{equation}	
where each projector is defined through an orthogonality condition with the corresponding tensor such that $\sum_{pol} \mathcal{P}_i T_j= \delta_{ij}$.  Since the resulting projectors are the same as those appearing in the literature~\cite{Peraro:2019cjj,Peraro:2020sfm} we do not list them explicitly here. 

Once the projectors are obtained, they can be applied to all relevant Feynman Diagrams to extract the corresponding helicity amplitude for a specific helicity configuration $\lambda_i$. We order the helicities as follows:
\begin{align}
	\label{helicities}
	&\lambda_1 = \{+,+,+,+\}, &&\lambda_2 = \{-,+,+,+\}, \\\nonumber
	&\lambda_3 = \{+,-,+,+\}, &&\lambda_4 = \{+,+,-,+\}, \\\nonumber
	&\lambda_5 = \{+,+,+,-\}, &&\lambda_6 = \{-,-,+,+\}, \\\nonumber
	&\lambda_7 = \{-,+,-,+\}, &&\lambda_8 = \{+,-,-,+\}. 
\end{align}	
The helicity amplitude $\mathcal{A}_{\lambda_j}$ for a specific configuration $\lambda_j$ is obtained by fixing the polarization vectors which occur in each $T_i$, which we write as $T_i^{\lambda_j}$:
\begin{equation}
	\mathcal{A}_{\lambda_j} = \sum_{i=1}^{8}\mathcal{F}_i \: T_{i}^{\lambda_j}.
\end{equation}
To lighten up the notation, we can further factorize out the overall spinor phases:
\begin{equation}
	\mathcal{A}_{\lambda_i} = \mathrm{S}_{\lambda_i}\mathrm{A}_{\lambda_i},
\end{equation}	
where the spinor phases $\mathrm{S}$ are given in the spinor-helicity formalism by:
\begin{align}
	&\mathrm{S}_{\lambda_1}  = \frac{[12][34]}{\langle1 2 \rangle\langle3 4 \rangle} , &&\mathrm{S}_{\lambda_2} = \frac{\langle12 \rangle\langle14 \rangle[24]}{\langle34 \rangle\langle23 \rangle\langle 24\rangle}, \\
	&\mathrm{S}_{\lambda_3}  = \frac{\langle21 \rangle\langle24 \rangle[14]}{\langle34 \rangle\langle13 \rangle\langle14 \rangle}, &&\mathrm{S}_{\lambda_4}  =  \frac{\langle32 \rangle\langle34 \rangle[24]}{\langle 14\rangle\langle12 \rangle\langle42 \rangle},\\
	&\mathrm{S}_{\lambda_5}  = \frac{\langle42 \rangle\langle43 \rangle[23]}{\langle13 \rangle\langle21 \rangle\langle23 \rangle}, &&\mathrm{S}_{\lambda_6}  =  \frac{\langle 12\rangle[34]}{[12]\langle34 \rangle},\\
	&\mathrm{S}_{\lambda_7}  = \frac{\langle13 \rangle[24]}{[13]\langle24 \rangle}, &&\mathrm{S}_{\lambda_8}  = \frac{\langle23 \rangle[14]}{[23]\langle14 \rangle},
\end{align}	
and the remaining amplitude coefficients are now purely scalar functions and are written as combination of form factors:
\begin{align}
	\label{FF}
	\mathrm{A}_{\lambda_1} = & \frac{t^2}{4}\left(\frac{2\mathcal{F}_{6}}{u}-\frac{2\mathcal{F}_{3}}{s}-\mathcal{F}_{1}  \right)+\mathcal{F}_{8}\left(\frac{s}{u}+\frac{u}{s}+4  \right)+\frac{t}{2}\left(\mathcal{F}_2-\mathcal{F}_4+\mathcal{F}_5-\mathcal{F}_7 \right), \\
	\mathrm{A}_{\lambda_2} = &  \frac{t^2}{4}\left(\frac{2\mathcal{F}_3}{s}+\mathcal{F}_1 \right) +t\left( \frac{\mathcal{F}_8}{s}+\frac{1}{2}\left(\mathcal{F}_4+\mathcal{F}_6-\mathcal{F}_2\right)  \right),\\
	\mathrm{A}_{\lambda_3} = &  -\frac{t^2}{4}\left(\frac{2\mathcal{F}_6}{u}-\mathcal{F}_1 \right)+t\left(\frac{\mathcal{F}_8}{s}+\frac{1}{2}\left(\mathcal{F}_2+\mathcal{F}_3+\mathcal{F}_5 \right) \right), \\
	\mathrm{A}_{\lambda_4} = &  \frac{t^2}{4}\left(\frac{2\mathcal{F}_3}{s}+\mathcal{F}_1 \right)+t\left(\frac{\mathcal{F}_8}{s}+\frac{1}{2}\left(\mathcal{F}_6+\mathcal{F}_7-\mathcal{F}_5 \right) \right) ,\\
	\mathrm{A}_{\lambda_5} = &  -\frac{t^2}{4}\left(\frac{2\mathcal{F}_6}{u}-\mathcal{F}_1 \right)+t\left(\frac{\mathcal{F}_8}{u}+\frac{1}{2}\left(\mathcal{F}_4+\mathcal{F}_7-\mathcal{F}_3 \right) \right) ,\\
	\mathrm{A}_{\lambda_6} = & -\frac{t^2}{4}\mathcal{F}_1+\frac{t}{2}\left(\mathcal{F}_2+\mathcal{F}_3-\mathcal{F}_6-\mathcal{F}_7 \right)+2\mathcal{F}_8, \\
	\mathrm{A}_{\lambda_7} = & t^2\left(\frac{\mathcal{F}_8}{su}-\frac{\mathcal{F}_3}{2s}+\frac{\mathcal{F}_6}{2u}-\frac{\mathcal{F}_1}{4}  \right) , \\
	\mathrm{A}_{\lambda_8} = & -\frac{t^2}{4}\mathcal{F}_1+\frac{t}{2}\left(\mathcal{F}_3-\mathcal{F}_4+\mathcal{F}_5-\mathcal{F}_6 \right)+2\mathcal{F}_8.
\end{align}

Since the interest of this paper is EW corrections, we will frequently encounter $\gamma_5$. We define $\gamma_5$ in the Larin prescription~\cite{Larin:1993tp,Larin:1993tq}, specifically:
\begin{equation}
	\label{larin}
	\gamma_5 = \frac{i}{24}\epsilon_{\mu\nu\rho\sigma}\gamma^{\mu}\gamma^{\nu}\gamma^{\rho}\gamma^{\sigma}.
\end{equation}
The Larin scheme requires a finite renormalization of $\gamma_5$ to ensure the validity of the Ward identity. However our calculation has no dependence on $\gamma_5$ at lowest order. This implies that through $\mathcal{O}(\epsilon^0)$ there should be no difference in the renormalized amplitude between our Larin choice above and a naive anti-commuting $\gamma_5$. We have checked this is indeed the case. The attentive reader may be concerned that the construction of our form factors described above utilized only CP even building blocks, and the introduction of EW corrections and $\gamma_5$ may introduce terms that depend on the CP-odd Levi-Civita tensor $\epsilon_{\mu\nu\rho\sigma}$.  First we note that any such terms may only appear linearly, since two Levi-Civita tensors can always be contracted as follows: 
\begin{eqnarray}
	\epsilon_{\mu_1\mu_2\mu_3\mu_4}\epsilon^{\nu_1\nu_2\nu_3\nu_4}= 
	\begin{vmatrix}
		&  \delta^{\nu_1}_{\mu_1}& \delta^{\nu_2}_{\mu_1} &\delta^{\nu_3}_{\mu_1} &\delta^{\nu_4}_{\mu_1} \\
		&  \delta^{\nu_1}_{\mu_2}& \delta^{\nu_2}_{\mu_2} &\delta^{\nu_3}_{\mu_2} &\delta^{\nu_4}_{\mu_2} \\
		&  \delta^{\nu_1}_{\mu_3}& \delta^{\nu_2}_{\mu_3} &\delta^{\nu_3}_{\mu_3} &\delta^{\nu_4}_{\mu_3} \\
		&  \delta^{\nu_1}_{\mu_4}& \delta^{\nu_2}_{\mu_4} &\delta^{\nu_3}_{\mu_4} &\delta^{\nu_4}_{\mu_4}
	\end{vmatrix},
\end{eqnarray}
and the RHS of this equation is clearly expressible in-terms of our pre-existing tensor structures. Secondly we note that a term linear in $\epsilon_{\mu\nu\rho\sigma}$ is orthogonal to all of the structures which enter into the LO projection when the polarization sum is performed. As a result, no mixed CP-odd/even can survive when we ultimately construct our desired interference with the LO amplitude. We conclude that excluding the $\epsilon_{\mu\nu\rho\sigma}$ term from our tensor structures could only possibly affect the square of the two-loop amplitude (i.e. the $\mathcal{O}(\alpha^2)$ contribution) which is beyond the scope of our work here. We therefore work with the eight projectors described above. 

\subsection{EW corrections to $gg\rightarrow \gamma\gamma$}

We are interested in the EW corrections to the $gg\rightarrow \gamma\gamma$ process. We will work in the Feynman gauge. Given that the $W$ bosons possess an electromagnetic charge, new structures enter the calculation at two-loops.  
We present the types of topology which enter our calculation in Fig.~\ref{fig:feynAll}.
\begin{figure}[h]
	\centering
	\begin{subfigure}{0.32\linewidth}
		\includegraphics[width=\linewidth]{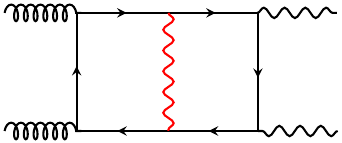}
		\caption{Topology P$_\text{I}$}
		\label{fig:PI_feyn}
	\end{subfigure}\hspace{0.1cm}
	\begin{subfigure}{0.32\linewidth}
		\includegraphics[width=\linewidth]{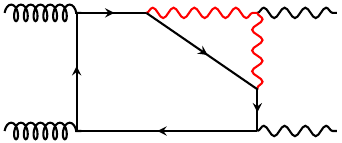}
		\caption{Topology P$_\text{II}$}
		\label{fig:PII_feyn}
	\end{subfigure}\hspace{0.1cm}
	\begin{subfigure}{0.32\linewidth}
		\includegraphics[width=\linewidth]{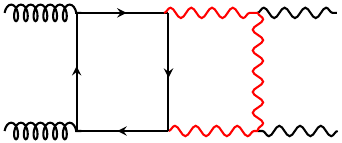}
		\caption{Topology P$_\text{III}$}
		\label{fig:PIII_feyn}
	\end{subfigure}\vspace{0.5cm}
	
	\begin{subfigure}{0.32\linewidth}
		\includegraphics[width=\linewidth]{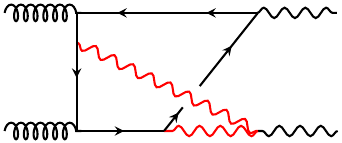}
		\caption{Topology N$_\text{I}$}
		\label{fig:NI_feyn}
	\end{subfigure}\hspace{0.5cm}
	\begin{subfigure}{0.32\linewidth}
		\includegraphics[width=\linewidth]{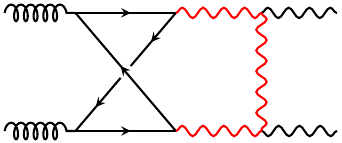}
		\caption{Topology N$_\text{II}$}
		\label{fig:NII_feyn}
	\end{subfigure}
	\caption{Feynman diagrams associated with  the relevant W-type topologies. Propagators marked in red are massive. }
	\label{fig:feynAll}
\end{figure}
There are three planar topologies and two non-planar topologies. The first topology P$_\text{I}$ is a correction to the lowest order diagram where a boson is exchanged between the closed loop of fermions. This topology therefore has contributions where the boson is either a photon, $Z$ or $W$ boson. Since we consider massless quarks, no Higgs or Goldstone contributions occur in this topology. The remaining topologies include multiple exchanges of $W$, and/or associated $\phi^{\pm}$ Goldstone bosons. We note that in addition to the three-point vertex illustrated in Fig.~\ref{fig:feynAll}, quartic $WW\gamma\gamma$ vertices also contribute in the $\text{N}_\text{II}$ and $\text{P}_\text{III}$ topologies. 

The bare matrix element for the process Eq.~\ref{eq:process} can be written as a perturbation series as follows:
\begin{equation}
	{\mathcal{A}}_{gg\to\gamma\gamma} = \delta^{ab}(4\alpha_0\alpha_{s}) \left[{\mathcal{A}}_{gg\to\gamma\gamma}^{(1)}+\frac{\alpha_0}{2\pi}{\mathcal{A}}_{gg\to\gamma\gamma}^{(2)}+\mathcal{O}(\alpha_0^2) \right],
\end{equation}
where ${\mathcal{A}}_{gg\to\gamma\gamma}^{(L)}$ defines the $L$-loop bare amplitude, and $\alpha_s$ and $\alpha_0$ are the bare QCD and EW couplings. The factor $\delta^{ab}$ defines the overall color factor. 
We will shortly discuss the UV renormalization of the process, in doing so it will be useful to introduce the charge-striped one-loop amplitude: 
\begin{eqnarray}
	\label{LO2}
	{\mathcal{A}}_{gg\to\gamma\gamma}^{(1)} = \left(\sum Q_f^2\right) \;\mathcal{C}^{(1)}_f,
\end{eqnarray}
where $\mathcal{C}^{(1)}_f$ is the combination of all one-loop box diagrams from a single fermion flavor, and the sum runs over all active flavors. 
It will be convenient to further split the 2-loop bare amplitude into different contributions depending on the nature of the exchanged boson:
\begin{equation}
	\label{amp_topo}
	{\mathcal{A}}_{gg\to\gamma\gamma}^{(2)} = {\mathcal{A}}_{\text{QED}}^{(2)}+{\mathcal{A}}_{Z}^{(2)}+{\mathcal{A}}_{W}^{(2)},
\end{equation}
where ${\mathcal{A}}_{\text{QED}}^{(2)}$ represents the pure QED contributions, ${\mathcal{A}}_{Z}^{(2)}$ are the $Z$ boson corrections and the ${\mathcal{A}}_{W}^{(2)}$ are the contributions coming from the $W$ (and $\phi$) diagrams. For the two-loop ${\mathcal{A}}_{\text{QED}}^{(2)}$ amplitude the charge completely factorizes, and we can decompose it as follows:
\begin{equation}
	\label{QED2}
	\mathcal{A}_{QED}^{(2)} = \left( \sum Q_f^4\right) \;\mathcal{C}^{(2)}_{\gamma},
\end{equation}
where the sum runs over the active flavors. For the ${\mathcal{A}}_{Z}^{(2)}$ amplitude, the up and down type contributions get different factors arising from the different choices of $v_f$ and $a_f$ in the $Z$-coupling:
\begin{equation}
	\label{Z2}
	\mathcal{A}_{Z}^{(2)} = \left[ \sum Q_f^2\left(v_{f}^2+a_{f}^2 \right) \right] \;\mathcal{C}^{(2)}_{Z},
\end{equation}
where the sum runs over the active flavors, and we recall that:
\begin{equation}
	v_f = \frac{I^3_{f}-2s_W^2Q_f}{2s_Wc_W} \qquad a_f =\frac{I_f^3}{2s_Wc_W} ,
\end{equation}
with $I_f^3$ being the weak isospin for the flavor $f$. Finally for ${\mathcal{A}}_{W}^{(2)}$ there are several different charge combinations based on whether the photon couples to zero, one or two of the fermions in the amplitude. For each fermionic generation, we decompose it as:
\begin{equation}
	\label{W2}
	\mathcal{A}^{(2)}_{W} = \sum\frac{1}{s_W^2} \left[\mathcal{C}_{W}^{0}+\mathcal{C}_{W}^{Q_uQ_d}Q_u Q_d+\mathcal{C}_{W}^{Q_u - Q_d}(Q_u - Q_d)+\mathcal{C}_{W}^{(Q_u - Q_d)^2}(Q_u - Q_d)^2\right],
\end{equation}
where, this time, the sum runs over the active fermion generations. 

We now define the spin- and color-averaged squared matrix element as
\begin{equation}
	\label{eq1}
	{\mathcal{M}}_{gg\to\gamma\gamma}(s,t,u) \equiv \frac{1}{N_{\text{stat}}N_{\text{spin}} \, N_{\text{color}}}
	\sum_{\text{spins, colors}} |{\mathcal{A}}_{gg\to\gamma\gamma}|^2,
\end{equation}
where $N_{\text{spin}}$ and $N_{\text{color}}$ represent the average initial state gluon spin and color and $N_{\text{stat}}=2$ is the statistical factor associated with the two identical final state photons.
Expanding Eq.~\ref{eq1} including $\mathcal{O}(\alpha)$ corrections we obtain:
\begin{eqnarray}
	\label{bare_amp1}
	{\mathcal{M}}_{gg\to\gamma\gamma}(s,t,u) = \frac{4\alpha_0^2\alpha_s^2}{(N_c^2-1)}\left[\mathcal{M}^{(1)}_{gg\to\gamma\gamma}+\frac{\alpha_0}{2\pi}{\mathcal{M}}^{(2)}_{gg\to\gamma\gamma} +\mathcal{O}(\alpha_0^2) \right],
\end{eqnarray}
where, by expanding the matrix element squared in-terms of the exchanged bosons as written above, we defined:
\begin{eqnarray}
	&&\label{piece1}
	\mathcal{M}^{(1)}_{gg\to\gamma\gamma} = \left| \mathcal{A}^{(1)}_{gg\to\gamma\gamma}\right|^2 ,\\
	\label{piece2}
	&& {\mathcal{M}}_{{X}}^{(2)} = 2\,\text{Re}\left[\left({\mathcal{A}}_{gg\to\gamma\gamma}^{(1) }\right)^*\hspace{-0.3em}\cdot{\mathcal{A}}_{{X}}^{(2)}\right] ,
\end{eqnarray}
with: 
\begin{eqnarray}
	{\mathcal{M}}^{(2)}_{gg\to\gamma\gamma}= {\mathcal{M}}_{{\text{QED}}}^{(2)}+ {\mathcal{M}}_{{{W}}}^{(2)}+ {\mathcal{M}}_{{{Z}}}^{(2)}.
\end{eqnarray}

\subsection{UV renormalization}
\label{sec:UVRen}

In order to express the bare two-loop amplitude defined in the previous section in a physically meaningful manner, we must renormalize couplings, wavefunctions and masses through $\mathcal{O}(\alpha).$ We note that the strong-coupling does not require renormalization at this order. 
To renormalize the electroweak sector, we work in the {on-shell (OS) scheme}. 
The explicit formulation of the OS-scheme and expressions for the subsequent counter-terms can be found in, for instance, refs.~\cite{Denner:2019vbn,Denner:1991kt}. 

Since our process contains a closed loop of fermions, the dependence on the fermion field strength cancel when the propagator and vertex corrections are combined, leaving only the terms arising from charge renormalization $(\delta Z_e)$ and photon wavefunction renormalization ($\delta Z_{AA}$). The total counter-term simplifies since they occur in the combination $\delta Z_e + \frac{1}{2}\delta Z_{AA}$, which can be replaced via Ward identities as follows:
\begin{equation}
	\label{ward_id}
	\delta Z_e + \frac{1}{2}\delta Z_{AA} = \frac{s_W}{c_W}\frac{1}{2}\delta Z_{ZA},
\end{equation}
where $\delta Z_{ZA}$ is expanded through $\epsilon^0$ as follows:  
\begin{equation}
	\delta Z_{ZA} =\frac{1}{\epsilon} \frac{\alpha}{\pi}\frac{c_W}{s_W } - \frac{\alpha}{\pi}\frac{c_W}{s_W} \log\left(\frac{M_W^2}{\mu^2} \right) + \mathcal{O}(\epsilon),
\end{equation}
where we observe that, compared to the convention in ref.~\cite{Denner:2019vbn}, the extra factor of $1/2$ comes from the different expansion parameter choice of $\alpha/(2\pi)$.

Summing over a single generation of up and down type quarks we find the following counter-term amplitude : 
\begin{eqnarray}
	\label{CT_All}
	\mathcal{A}_{\text{CT}} = \left[ \frac{(Q_u-Q_d) }{4c_W s_W} \delta Z_{ZA}\right] \mathcal{C}^{(1)}_{q},
\end{eqnarray}
such that the total contribution to the matrix-element squared for two generations is:
\begin{eqnarray}
	\mathcal{M}_{\text{CT}}  = 2 \text{Re}\left[\left({\mathcal{A}}_{gg\to\gamma\gamma}^{(1) }\right)^*\hspace{-0.3em}\cdot{\mathcal{A}}_{{CT}}\right].  
\end{eqnarray}
Expanding through $\mathcal{O}(\epsilon^0)$: 
\begin{eqnarray}
	\mathcal{M}_{\text{CT}} = \left(\mathcal{M}_{\text{CT}}^{\text{pole}} +\mathcal{M}_{\text{CT}}^{\text{fin}} \right) \mathcal{M}^{(1)}_{gg\to\gamma\gamma},
\end{eqnarray}
with
\begin{equation}
	\mathcal{M}_{\text{CT}}^{\text{pole}} = \left(\frac{1}{\epsilon}\right)\frac{  (Q_u-Q_d)^2}{2s_W^2}, \qquad \qquad \mathcal{M}_{\text{CT}}^{\text{fin}} = \frac{\left(Q_u-Q_d \right) }{2s_W^2}\log{\frac{\mu^2}{M_W^2}}.
\end{equation}
Finally, including all of the various diagrams and counter-term contributions, the UV renormalized $\mathcal{O}(\alpha)$ correction to the matrix-element squared can be written as:
\begin{equation}
	\label{amp_topo2}
	\tilde{\mathcal{M}}_{gg\to\gamma\gamma}^{(2)} = {\mathcal{M}}_{\text{QED}}^{(2)} + {\mathcal{M}}_{Z}^{(2)} + {\mathcal{M}}_{W}^{(2)}+ {\mathcal{M}}_{\text{CT}}.
\end{equation}
We observe that $\tilde{\mathcal{M}}_{gg\to\gamma\gamma}^{(2)}$ is finite as $\epsilon \rightarrow 0$. 

In order to evaluate the matrix element squared we must provide a scheme to determine $\alpha$. We note that, since photons are present in the final state, which have $Q^2=0$ on-shell, two of the three powers of $\alpha$ should be evaluated at zero momentum-transfer (discussed for instance, in ref.~\cite{Denner:2019vbn}). This leaves one remaining power of $\alpha$ at NLO, as well as two powers of the strong coupling. We choose to evaluate the remaining powers at a scale of $M_Z^2$, noting that the relationship between $\alpha(0)$ and $\alpha(M_Z^2)$ is as follows, 
\begin{equation}
	\alpha(M_Z^2) = \frac{1}{1-\Delta\alpha(M_Z^2)}\alpha(0),
\end{equation}
and 
\begin{equation}
	\Delta\alpha(M_Z^2) = \Delta\alpha_{\text{had}}^{(5)}(M_Z^2) + \Delta\alpha_{\text{lep}}(M_Z^2) + \Delta\alpha_{\text{top}}(M_Z^2).
\end{equation}
We see that through $\mathcal{O}(\alpha^3)$ the mixed-scale scheme and pure $\alpha(0)$ schemes are equivalent. However, since the remaining power of $\alpha$ arises from internal degrees of freedom, it is susceptible to large corrections in the $\alpha(0)$-scheme brought on by vacuum polarization loops and light quark masses. The use of $\alpha(M_z^2)$ mitigates the dependence on light quark masses (at the next order) and in essence captures some of the higher order corrections beyond the scope of our calculation. 


\section{Amplitude Evaluation}
\label{sec:MIeval}

\subsection{Amplitude Assembly and Validation}

In the previous section we defined the two-loop amplitude in terms of a set of eight form-factors, here we discuss the calculation and evaluation of the EW corrections to the form-factors. Our calculation begins by generating all the relevant Feynman diagrams using \texttt{Qgraf}, excluding the third generation of fermions. Next we use an in-house \texttt{Form} package to implement the Feynman rules and subsequent Dirac and color algebra evaluation. We further simplified the result using spinor-helicity properties, resulting in expressions for each of the eight scalar form-factors at the desired order. 

At this stage the scalar form-factors contain thousands of loop integrals, which must be reduced to a basis of Master Integrals (MI's). 
To perform the reduction, we use \texttt{Kira}~\cite{Klappert:2020nbg} to generate the IBP identities needed to relate the scalar integrals appearing in the helicity amplitudes in terms of the chosen Laporta basis. These IBP systems, particularly for non-planar sectors, are extremely large and can introduce a large amount of redundancy in the analytic expressions. Therefore, a dedicated cleanup step is essential to manage both complexity and size for the following steps. This simplification is carried out in \texttt{Mathematica}, using in-house routines developed specifically to streamline the application of the IBPs and clean-up the helicity amplitudes at this preliminary stage. 

At this stage each integral topology is written in terms of the Laporta basis master integrals. In order to make the MI's more amenable to calculation we next transform the basis to a canonical one, as outlined in our companion paper~\cite{Fiore:2023myh}. We refer the interested reader to this paper and here we present a brief summary of that work relevant for this section. We recall that there are five topologies to calculate, and  of the five topologies three are planar and two are non-planar.  A summary of the five topologies is presented in Fig.~\ref{fig:feynAll}. We note that for the most complicated topologies we expressed many of the MIs in terms of Chen-iterated integrals. These integrals require a two-parameter integral for the evaluation of a weight-4 contribution. For the topologies which contain exclusively integrals which can be expressed in-terms of Goncharov Polylogarithms (GPLs), there are thousands of unique weight-4 GPLs to evaluate. 

After expanding our canonical basis integrals (and associated coefficients from the reduction) in $\epsilon$ we finally arrive at an $\epsilon$ series for our amplitude.
At this stage we are ready to perform the UV-renormalization outlined in the previous section, ultimately arriving at our final result. 

Given its overall complexity, we performed extensive testing on our results to ensure their correctness. Firstly, in order to test the implementation of our results from ref.~\cite{Fiore:2023myh} we calculated each integral which appeared in the initial form factor (before application of the IBPs) numerically using \texttt{AMFlow}~\cite{Liu:2022chg} at a couple of reference points. This allowed us to calculate the amplitude without using any of the steps from the \texttt{Kira} stage on-wards and allowed us to verify and validate our analytic calculation of the MIs, implementation, and simplification of the IBP results. 

Next, we recall that at our order the $gg\rightarrow \gamma\gamma$ process is infrared (IR) finite. The finiteness of the total amplitude relies on non-trivial cancellations between Feynman diagrams with differing permutations of the photon and gluon orderings (i.e. under various exchanges of Mandelstam invariants). This can readily be seen even at one-loop, any given box diagram contains IR poles at $\mathcal{O}(\epsilon^{-2})$ arising from the zero-mass box integral. However, when the full set of 6 different box diagrams are summed the individual IR terms cancel and the one-loop amplitude is finite. For our two-loop calculation the cancellation of IR poles is more intricate. For instance in the QED and $Z$-type corrections the IR poles begin at fourth order $\epsilon^{-4}$, the cancellations therefore occur across four different terms in the $\epsilon$-expansion with increasing complexity at each order. The $W$-type diagrams are even more intricate and involve delicate cancellations across different charge structures and loop integral topologies. We have performed this calculation analytically, ensuring finiteness of our (UV-renormalized) result. Given the detailed and non-trivial cancellations described above this provides a robust test on the correctness of our calculation. 

In order to further test the finite part ($\epsilon^0$) of our calculation we performed several additional checks. Firstly we note that $\gamma_5$ first enters the calculation at this order, therefore, the finite renormalization of $\gamma_5$ required in the Larin scheme to ensure correctness of the Ward Identity plays no part. In other words the Larin scheme and a naive four-dimensional implementation of $\gamma_5$ should yield the same results to $\mathcal{O}(\epsilon^0)$ if used consistently across the calculation. We tested this by calculating the amplitude using both setups, finding perfect agreement. A further useful check of the result comes from the Bose symmetry of the process under exchange of the final state photons ($t \leftrightarrow u$). We have performed this check, again finding perfect agreement. Finally, the one-loop amplitude and two-loop contributions from QED have been validated against the literature, finding perfect agreement~\cite{Bern:2001df,Bern:2002jx}.

Having extensively checked and validated our calculation as described above we present our amplitude evaluated at a test phase space point, we define:
\begin{equation}
	x = - \frac{s}{M_W^2},\qquad y = -\frac{t}{M_W^2}, 
\end{equation}
then we choose $x = -3.80115$ and $y = 1.91294$ as a representative point. We further break the results down by helicity and mediating bosons as in Eq.~\ref{amp_topo}. Table~\ref{table:amplitudes} contains the numerically evaluated $\epsilon^0$ term for the one-loop, QED and $Z$-mediated bare amplitude coefficients, as defined in Eq.~\ref{LO2},~\ref{QED2} and \ref{Z2} respectively. The corresponding helicity configuration can be read from Eq.~\ref{helicities}.
For the $W$ topologies, we further decompose the two-loop amplitude as described in Section~\ref{sec:overview} according to their charge structure. The corresponding numerical coefficients are reported in Table~\ref{table:amplitudes_W} for each helicity configuration.
\begin{table}[H]
	\scriptsize
	\centering
	\begin{tabular}{|c|c c c|}
		\hline
		\rule{0pt}{10pt}
		Helicity & $\mathcal{C}^{(1)}_{f}$  & $\mathcal{C}^{(2)}_{\gamma}$  & $\mathcal{C}^{(2)}_{Z}$ \\
		\hline
		\rule{0pt}{12pt}
		$\lambda_1$ & 1.00000 & -3.00000 & -0.160732 + 1.14943 $i$ \\[3pt]
		$\lambda_2$ & 1.00000 & 0.35544 - 1.46320 $i$ & -0.271924 - 0.625633 $i$ \\[3pt]
		$\lambda_3$ & 1.00000 & 0.35544 - 1.46320 $i$ & -0.271924 - 0.625633 $i$ \\[3pt]
		$\lambda_4$ & 1.00000 & 0.35544 - 1.46320 $i$ & -0.271924 - 0.625633 $i$ \\[3pt]
		$\lambda_5$ & 1.00000 & 0.35544 - 1.46320 $i$ &  -0.271924 - 0.625633 $i$\\[3pt]
		$\lambda_6$ & -3.46746 & 2.48148 + 6.32858 $i$ &  2.68025 + 2.13708 $i$ \\[3pt]
		$\lambda_7$ & -0.11940 + 1.44926$i$ & 0.03633 + 2.3001 $i$ & 0.355133 + 1.10661 $i$ \\[3pt]
		$\lambda_8$ & -0.12402 + 1.47714$i$ & 0.02125 + 2.3295 $i$ & 0.351172 + 1.13133 $i$\\[3pt]
		\hline 
	\end{tabular}
	\caption{Numerical evaluation of $\epsilon^0$ part of the bare 1-loop and the 2-loop QED and $Z$ coefficients for various helicity configurations, as defined in Eq.~\ref{LO2},~\ref{QED2} and \ref{Z2} respectively. The kinematic point corresponds to $x = -3.80115$ and $y = 1.91294$ and the helicities $\lambda_i$ are defined in Eq.~\ref{helicities}.} 
	\label{table:amplitudes}
\end{table}

\begin{table}[H]
	\scriptsize
	\centering
	\begin{tabular}{|c|c c c c|}
		\hline
		\rule{0pt}{10pt}
		Helicity & $\mathcal{C}_{\lambda_i}^{0}$ & $\mathcal{C}_{W}^{Q_u Q_d}$ & $\mathcal{C}_{W}^{Q_u - Q_d}$ & $\mathcal{C}_{W}^{(Q_u - Q_d)^2}$ \\
		\hline
		\rule{0pt}{12pt}
		$\lambda_1$ & 0.89257 - 0.82938 $i$ & -0.08037 + 0.57472 $i$ & -0.62048 + 0.60892 $i$ & 0.19855 - 0.16164 $i$ \\[3pt]
		$\lambda_2$ & 0.03308 - 0.01780 $i$ & -0.13596 - 0.31282 $i$ & 2.00279 + 0.57157 $i$ & -0.56054 - 0.10370 $i$ \\[3pt]
		$\lambda_3$ & 0.26818 + 0.68678 $i$ & -0.13596 - 0.31282 $i$ & 1.72811 - 0.32802 $i$ & -0.52096 + 0.09131 $i$ \\[3pt]
		$\lambda_4$ & 0.22396 + 0.22457 $i$ & -0.13596 - 0.31282 $i$ & 1.92142 + 0.60820 $i$ & -0.56427 - 0.16371 $i$ \\[3pt]
		$\lambda_5$ & 0.22387 + 0.22454 $i$ & -0.13596 - 0.31282 $i$ & 1.92131 + 0.60773 $i$ & -0.56407 - 0.16321 $i$ \\[3pt]
		$\lambda_6$ & -4.02877 - 1.82061 $i$ & 1.34012 + 1.06854 $i$ & -1.42633 - 6.53959 $i$ & -0.04268 + 2.16402 $i$ \\[3pt]
		$\lambda_7$ & -0.35158 + 0.81397 $i$ & 0.17757 + 0.55378 $i$ & -0.60765 + 0.87088 $i$ & 0.12463 - 0.13865 $i$ \\[3pt]
		$\lambda_8$ & -0.12284 + 1.54886 $i$ & 0.17559 + 0.56567 $i$ & -0.90004 - 0.03496 $i$ & 0.16580 + 0.05991 $i$ \\[3pt]
		\hline 
	\end{tabular}
	\caption{Numerical evaluation of the $\epsilon^0$ part of the coefficients defined in Eq.~\eqref{W2} for the 2-loop $W$-type contributions to the bare amplitude. The kinematic point is $x = -3.80115$ and $y = 1.91294$, and the helicities are defined in Eq.~\ref{helicities}. $\phi$ Goldstone bosons are also included with the $W$ diagrams in these results.}  
	\label{table:amplitudes_W}
\end{table}

\subsection{Master Integral Evaluation for Phenomenology}

As discussed in the previous section the calculation of our amplitude requires the evaluation of MI's for five distinct topologies (and additionally those arising from exchange of a photon which we do not consider in this section since they are much simpler). The calculation of the MIs in a canonical basis results in thousands of weight 4 GPLs and two-fold Chen iterated integrals to evaluate. These features are undesirable when attempting to perform a phenomenological study in a MC environment, since hundred of thousands of phase space points are typically evaluated. One therefore needs to implement the MIs in such a way to ensure a reasonable computational evaluation time. In order to do this we utilize our analytic results to generate grids and use fitting functions to extrapolate fits to the MIs from said data. We describe this process in this section. 

Chen iterated integrals, which naturally exhibit logarithmic behavior are well suited for such an analytic representation using carefully chosen fitting functions. Such representations also provide a straightforward way to handle both physical and non-physical thresholds over the phase space. Our approach relies on the observation that the underlying master integrals possess logarithmic behavior that can be mimicked by a well-chosen rational polynomial function. To this end, we opt to fit the real and imaginary parts of the integrals separately since, due to analytic continuation over different regions, they exhibit very different behavior. \\
Furthermore, we found it significantly more efficient and robust to fit the MI's individually, rather than attempting to fit the complete helicity amplitudes. This separation is crucial, as fitting the full amplitudes would lose the underlying logarithmic structure, making it substantially more difficult to construct appropriate fitting functions. Such an approach would likely result in a loss of precision and a high risk of over-fitting. In contrast, by fitting only the order-by-order expansion of the master integrals while retaining the analytic form of the pre-coefficients, we achieved high numerical precision and stability. Having an analytic expression for the MIs was particularly helpful since it allowed us to generate large grids efficiently for fitting. \\

For the mixed Chen/GPL topologies, specifically $\text{P}_{\text{III}}$ and $\text{N}_{\text{II}}$, our procedure begins with the extraction of the specific $\epsilon$ order that enters the amplitude for each integral. In cases where the lower-order contributions can be represented in terms of GPLs, no fitting is required. Table~\ref{fitmapPIII} and Table~\ref{fitmapNII} show, for topologies $\text{P}_{\text{III}}$ and $\text{N}_{\text{II}}$ respectively,  the relevant weights which have to be fitted, and those which can be written in terms of GPLs. We observe that, for a given MI, not all weights contribute to the total cross section. Since some of these integrals may appear repeatedly across different topologies, the contribution from some of their weights may vanish once summed, due to prefactor cancellations. The dashes in Tables~\ref{fitmapPIII} and~\ref{fitmapNII} reflect this behavior. 

In summary, we found that rational polynomials for the real and imaginary parts, in the kinematic variables $x$ and $y$, suffice to capture the behavior of the mixed representation integrals:		\begin{equation}
	\mathcal{K}_{n}^{(w)}(x,y) = \frac{ \sum_{i+j=l} a_{i,j} x^i y^j }{\sum_{i+j=m} b_{i,j} x^i y^j}+i \frac{ \sum_{i+j=p} c_{i,j} x^i y^j }{\sum_{i+j=q} d_{i,j} x^i y^j},
\end{equation}
where, for any given topology, $\mathcal{K}_{n}^{(w)}(x,y)$ denotes the fitting representation of the integral $n$ of weight $w$, and the coefficients $a_{i,j}$, $b_{i,j}$, $c_{i,j}$ and $d_{i,j}$ are determined through the fitting procedure. The degree of the polynomials is dictated through the parameters $\{l,m,p,q \}$. For most of the integrals belonging in a low-propagator family (typically up to 5-propagators) we found that an overall degree of 6 is sufficient to reach the desired precision for the fitting representation. However, integrals with 6 and 7 propagators which enter both topologies, such as integrals $\mathcal{G}_{30}^{\text{P}_{\text{III}}}$, $\mathcal{G}_{31}^{\text{P}_{\text{III}}}$ and $\mathcal{G}_{32}^{\text{P}_{\text{III}}}$ for $\text{P}_{\text{III}}$, are more complicated. In particular, those integrals have a higher number of pseudo-thresholds and have less evident logarithmic structure. For such cases we needed to employ polynomials of higher degree (up to 11th) to achieve the required precision.

For each topology several unique fitting regions are utilized, which we present in Fig~\ref{fitting_map}. This is to ensure our functions are reliable and do not cross pseudo-thresholds. Each colored region of Fig.~\ref{fitting_map} represents a region with a unique fit to the loop integrals within that topology.   We also note that, to validate our analytic results in each new region we checked our MI calculations with \texttt{AMFlow}.

Given the success of this methodology, we extended it beyond purely Chen-based topologies and applied it to the $\text{N}_\text{I}$ topology as well. While this topology can be fully expressed in terms of GPLs, its evaluation is computationally expensive due to the complexity of the 7-propagator family and the large number of weight-4 GPLs involved. By employing our fitting approach, we significantly reduced computational costs while maintaining accuracy comparable to that of the full analytic representation. Since both analytic and fitted representations are available, we implemented both  enabling a direct comparison of how fitting uncertainties propagate into a cross section.  Specifically, we calculated the integrated cross section arising only from the $\text{N}_\text{I}$ Feynman diagrams (i.e. an isolated component of the total cross section) using two versions of the bare helicity amplitude. The first version, $\sigma^{(\mathcal{K})}_{\text{N}_\text{I}}$, takes the relevant weight 3 and 4 integrals fitted using the procedure described above and the second one, $\sigma^{(\mathcal{G})}_{\text{N}_\text{I}}$, corresponds to the amplitude evaluated using the master integrals with a full GPL representation. We calculated the cross section in the region $90\,\text{GeV} < x < 150\,\text{GeV}$\footnote{Employing the phase space selection criteria defined in Section~\ref{sec:Results}} and evaluated the corresponding ratio between the two cross-sections. The results are presented in Table~\ref{tab:sigma_comparison}:
\begin{table}[H]
	\centering
	\begin{tabular}{ccc}
		\hline
		\textbf{Ratio $\sigma^{(\mathcal{K})}_{\text{N}_\text{I}}/\sigma^{(\mathcal{G})}_{\text{N}_\text{I}}$ } & \textbf{Uncertainty} \\
		\hline
		 $1.00001$       & $\pm\,0.007 $ \\
		\hline
	\end{tabular}
	\caption{Ratio between the two partial cross-section obtained using two different master integral implementations ($\mathcal{K}$ vs.\ $\mathcal{G}$). The cross-sections are obtained from the $\text{N}_\text{I}$ Topology only. }
	\label{tab:sigma_comparison}
\end{table}
We observe that the two partial cross sections agree, and that the fitting uncertainties are much smaller than the Monte Carlo integration uncertainties. This result gives us confidence in the fitting procedure for use in further phenomenological applications. 

\begin{figure}[H]
	\begin{subfigure}{0.32\linewidth}
		\centering
		\includegraphics[width=1\linewidth]{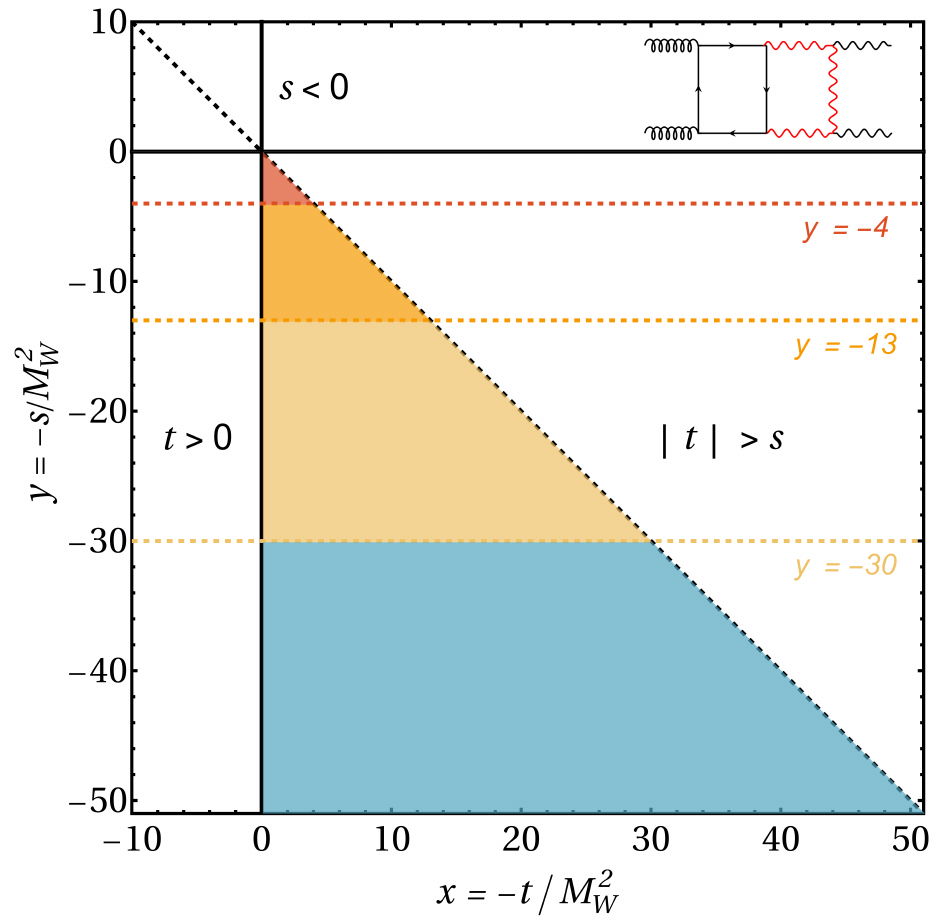}
	\end{subfigure}
	\begin{subfigure}{0.32\linewidth}
		\centering
		\includegraphics[width=1\linewidth]{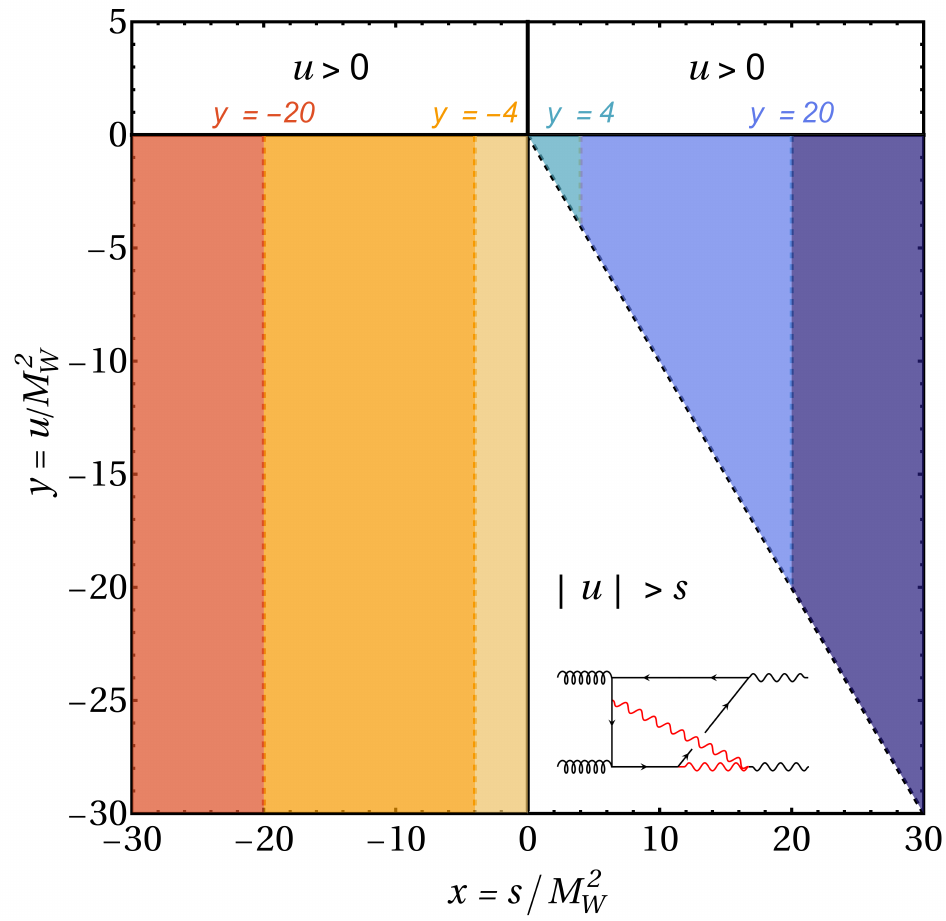}
	\end{subfigure}
	\begin{subfigure}{0.32\linewidth}
		\centering
		\includegraphics[width=1\linewidth]{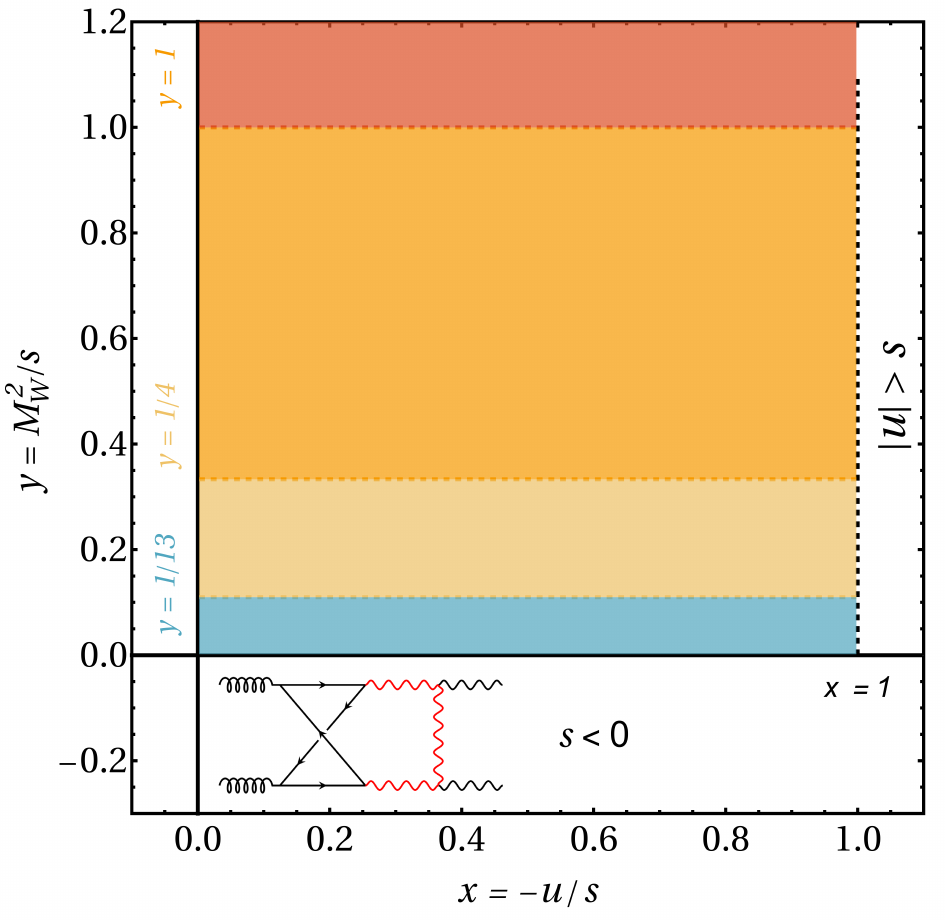}
	\end{subfigure}
	\caption{Break-down of the fitting region for each relevant topology, each color represents a unique fitting region.}
	\label{fitting_map}
\end{figure}

\begin{table}[H]
	\centering
	\renewcommand{\arraystretch}{1.2}
	\begin{tabular}{cc}
		\begin{tabular}{c|cccc}
			& $w=1$ & $w=2$ & $w=3$ & $w=4$ \\
			\hline
			$\mathcal{G}_{1}^{\text{P}_\text{III}}$ & \xmark & \xmark & \cmark & \cmark \\
			$\mathcal{G}_{3}^{\text{P}_\text{III}}$ & \xmark & \xmark & \cmark & \cmark \\
			$\mathcal{G}_{5}^{\text{P}_\text{III}}$ & \xmark & \xmark & \cmark & - \\
			$\mathcal{G}_{7}^{\text{P}_\text{III}}$ & \xmark & \xmark & \cmark & \cmark \\
			$\mathcal{G}_{9}^{\text{P}_\text{III}}$ & \xmark & \xmark & - & - \\
			$\mathcal{G}_{11}^{\text{P}_\text{III}}$ & \xmark & \xmark & - & - \\
			$\mathcal{G}_{13}^{\text{P}_\text{III}}$ & \xmark & \xmark & - & \cmark \\
			$\mathcal{G}_{15}^{\text{P}_\text{III}}$ & \xmark & \xmark & \cmark & \cmark \\
			$\mathcal{G}_{17}^{\text{P}_\text{III}}$ & \xmark & \xmark & \cmark & \cmark \\
			$\mathcal{G}_{19}^{\text{P}_\text{III}}$ & \xmark & \xmark & \cmark & \cmark \\
			$\mathcal{G}_{21}^{\text{P}_\text{III}}$ & \xmark & \xmark & \cmark & \cmark \\
			$\mathcal{G}_{23}^{\text{P}_\text{III}}$ & \xmark & \xmark & \cmark & -\\
			$\mathcal{G}_{25}^{\text{P}_\text{III}}$ & \xmark & \cmark & \cmark & -\\
			$\mathcal{G}_{27}^{\text{P}_\text{III}}$ & \xmark & \xmark & - & \cmark \\
			$\mathcal{G}_{29}^{\text{P}_\text{III}}$ & \xmark & \cmark & \cmark & -\\
			$\mathcal{G}_{31}^{\text{P}_\text{III}}$ & \xmark & \cmark & \cmark & \cmark \\
		\end{tabular}
		&
		\begin{tabular}{c|cccc}
			& $w=1$ & $w=2$ & $w=3$ & $w=4$ \\
			\hline
			$\mathcal{G}_{2}^{\text{P}_\text{III}}$ & \xmark & \xmark & \cmark & \cmark \\
			$\mathcal{G}_{4}^{\text{P}_\text{III}}$ & \xmark & \xmark & \cmark & \cmark \\
			$\mathcal{G}_{6}^{\text{P}_\text{III}}$ & \xmark & \xmark & \cmark & - \\
			$\mathcal{G}_{8}^{\text{P}_\text{III}}$ & \xmark & \xmark & \cmark & - \\
			$\mathcal{G}_{10}^{\text{P}_\text{III}}$ & \xmark & \xmark & - & - \\
			$\mathcal{G}_{12}^{\text{P}_\text{III}}$ & \xmark & \xmark & - & - \\
			$\mathcal{G}_{14}^{\text{P}_\text{III}}$ & \xmark & \xmark & - & \cmark \\
			$\mathcal{G}_{16}^{\text{P}_\text{III}}$ & \xmark & \xmark & \cmark & \cmark \\
			$\mathcal{G}_{18}^{\text{P}_\text{III}}$ & \xmark & \cmark & - & - \\
			$\mathcal{G}_{20}^{\text{P}_\text{III}}$ & \xmark & \xmark & \cmark & \cmark \\
			$\mathcal{G}_{22}^{\text{P}_\text{III}}$ & \xmark & \xmark & \cmark & - \\
			$\mathcal{G}_{24}^{\text{P}_\text{III}}$ & \xmark & \xmark & \cmark & \cmark \\
			$\mathcal{G}_{26}^{\text{P}_\text{III}}$ & \xmark & \xmark & \cmark & \cmark \\
			$\mathcal{G}_{28}^{\text{P}_\text{III}}$ & \xmark & \xmark & \cmark & - \\
			$\mathcal{G}_{30}^{\text{P}_\text{III}}$ & \xmark & \xmark & - & \cmark \\
			$\mathcal{G}_{32}^{\text{P}_\text{III}}$ & \xmark & \xmark & \cmark & \cmark \\
		\end{tabular}
	\end{tabular}
	\caption{Map of the inclusion of weights and integrals $\mathcal{G}_n^{(w)}$ for topology $\text{P}_\text{III}$, arranged row-wise in two columns for readability. A red cross (\xmark) and a dash denotes that the corresponding integral and weight were excluded from the fitting procedure, while a green check mark (\cmark) indicates inclusion.}
	\label{fitmapPIII}
\end{table}
\begin{table}[H]
	\centering
	\renewcommand{\arraystretch}{1.2}
	\begin{tabular}{c|cccc@{\hspace{1cm}}c|cccc}
		& $w=1$ & $w=2$ & $w=3$ & $w=4$ & & $w=1$ & $w=2$ & $w=3$ & $w=4$ \\
		\hline
		$\mathcal{G}_{1}^{\text{N}_\text{II}}$ & \xmark & \xmark & \cmark & - & $\mathcal{G}_{2}^{\text{N}_\text{II}}$ & \xmark & \xmark & \cmark & - \\
		$\mathcal{G}_{3}^{\text{N}_\text{II}}$ & \xmark & \xmark & \cmark & - & $\mathcal{G}_{4}^{\text{N}_\text{II}}$ & \xmark & \cmark & \cmark & - \\
		$\mathcal{G}_{5}^{\text{N}_\text{II}}$ & \xmark & \cmark & \cmark & - & $\mathcal{G}_{6}^{\text{N}_\text{II}}$ & \xmark & \cmark & \cmark & - \\
		$\mathcal{G}_{7}^{\text{N}_\text{II}}$ & \xmark & \cmark & \cmark & - & $\mathcal{G}_{8}^{\text{N}_\text{II}}$ & \xmark & \xmark & \cmark & - \\
		$\mathcal{G}_{9}^{\text{N}_\text{II}}$ & \xmark & \xmark & - & - & $\mathcal{G}_{10}^{\text{N}_\text{II}}$ & \xmark & \xmark & \cmark & - \\
		$\mathcal{G}_{11}^{\text{N}_\text{II}}$ & \xmark & \xmark & \cmark & - & $\mathcal{G}_{12}^{\text{N}_\text{II}}$ & \xmark & \xmark & \cmark & \cmark \\
		$\mathcal{G}_{13}^{\text{N}_\text{II}}$ & \xmark & \xmark & \cmark & - & $\mathcal{G}_{14}^{\text{N}_\text{II}}$ & \xmark & \xmark & \cmark & \cmark \\
		$\mathcal{G}_{15}^{\text{N}_\text{II}}$ & \xmark & \cmark & - & - & $\mathcal{G}_{16}^{\text{N}_\text{II}}$ & \xmark & \xmark & \cmark & - \\
		$\mathcal{G}_{17}^{\text{N}_\text{II}}$ & \xmark & \cmark & \cmark & - & $\mathcal{G}_{18}^{\text{N}_\text{II}}$ & \xmark & \xmark & \cmark & - \\
		$\mathcal{G}_{19}^{\text{N}_\text{II}}$ & \xmark & \xmark & \cmark & - & $\mathcal{G}_{20}^{\text{N}_\text{II}}$ & \xmark & \xmark & \cmark & - \\
		$\mathcal{G}_{21}^{\text{N}_\text{II}}$ & \xmark & \cmark & \cmark & - & $\mathcal{G}_{22}^{\text{N}_\text{II}}$ & \xmark & \cmark & \cmark & \cmark \\
		$\mathcal{G}_{23}^{\text{N}_\text{II}}$ & \xmark & \cmark & \cmark & - & $\mathcal{G}_{24}^{\text{N}_\text{II}}$ & \xmark & \xmark & \cmark & \cmark \\
		$\mathcal{G}_{25}^{\text{N}_\text{II}}$ & \xmark & \cmark & - & - & $\mathcal{G}_{26}^{\text{N}_\text{II}}$ & \xmark & \xmark & \cmark & - \\
		$\mathcal{G}_{27}^{\text{N}_\text{II}}$ & \xmark & \xmark & \cmark & - & $\mathcal{G}_{28}^{\text{N}_\text{II}}$ & \xmark & \cmark & - & - \\
		$\mathcal{G}_{29}^{\text{N}_\text{II}}$ & \xmark & \xmark & \cmark & - & $\mathcal{G}_{30}^{\text{N}_\text{II}}$ & \xmark & \cmark & \cmark & - \\
		$\mathcal{G}_{31}^{\text{N}_\text{II}}$ & \xmark & \xmark & \cmark & - & $\mathcal{G}_{32}^{\text{N}_\text{II}}$ & \xmark & \cmark & \cmark & \cmark \\
	\end{tabular}
	\caption{Map of the inclusion of weights and integrals $\mathcal{G}_n^{(w)}$ for topology $\text{N}_\text{II}$, arranged row-wise in two columns for readability. A red cross (\xmark) and a dash denotes that the corresponding integral and weight were excluded from the fitting procedure, while a green check mark (\cmark) indicates inclusion.}
	\label{fitmapNII}
\end{table}

\subsection{Fitting Uncertainty by Region}

In the following, we present the relative uncertainties associated with the fitting procedure for the 7-propagator integrals in the two Chen-based topologies, $\text{P}_{\text{III}}$ and $\text{N}_{\text{II}}$. These represent the more challenging integrals to fit so, as to avoid overwhelming the manuscript with repetitive figures, we will only present the results for these. It should be noted that for lower propagator integrals the fitting uncertainties are much smaller than those presented here.The error maps were generated by evaluating a grid of kinematic points using the Chen integral representation $\mathcal{G}$ and comparing the results with the corresponding fitted functions $\mathcal{K}$. As stated before, for certain weights some of the relevant integrals in these Chen topologies can still be expressed analytically in terms of GPLs, especially at lower $\epsilon$ orders. As a result, for integrals at weight 1 fitting was unnecessary. The accuracy of the weight 2 fits was also very good ($10^{-12}$), so for brevity we focus on weight 3 and weight 4 for the most challenging integrals. 

Figs.[\ref{PIII_err_B_31}-\ref{NII_err_C1_32}] show the distribution of the relative difference between fitting functions and Chen integrals. The plots have been made with 10,000 points randomly chosen in each sub-region, and to avoid repetition, we  present a selected sub-region for both topologies. The relative uncertainties are kept below $10^{-6}$ for most points in the sub-region, the only exception being around pseudo-thresholds in the dlog sub-alphabets or near the region physical thresholds corresponding to $s = M_W^2$ and $s = 4M_W^2$.  This is partly because the class of the fitting functions chosen are indeed less flexible in capturing steep gradients or non-analytic behaviors. Even for those more problematic points, however, the relative difference are largely maintained below $10^{-5}$, with the maximum discrepancy reaching $10^{-4}$ for points extremely close to the physical thresholds. 

\begin{figure}[H]
	\centering
	\begin{subfigure}{0.8\linewidth}
		\centering
		\includegraphics[width=1\linewidth]{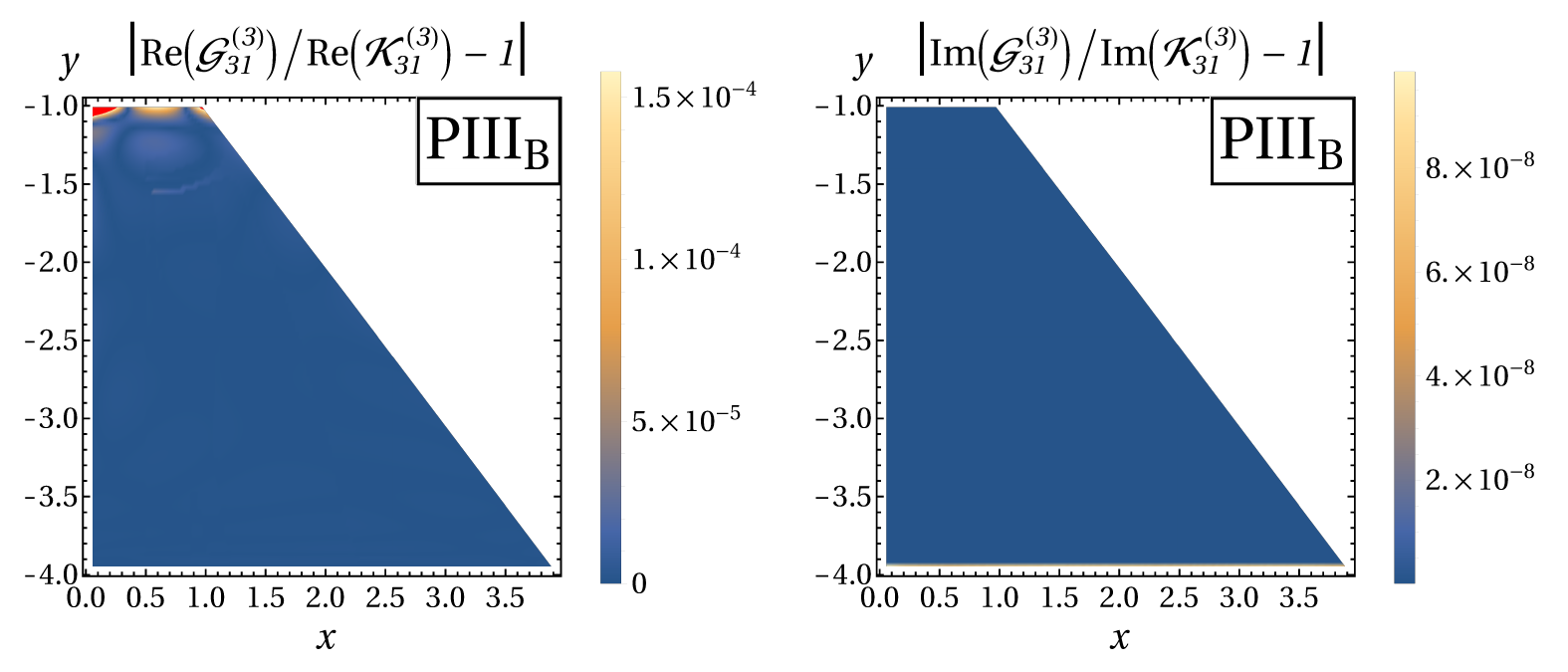}
		\label{PIII_err_B_31w3}
	\end{subfigure}\vspace{0.2cm}	
	\begin{subfigure}{0.8\linewidth}
		\centering
		\includegraphics[width=1\linewidth]{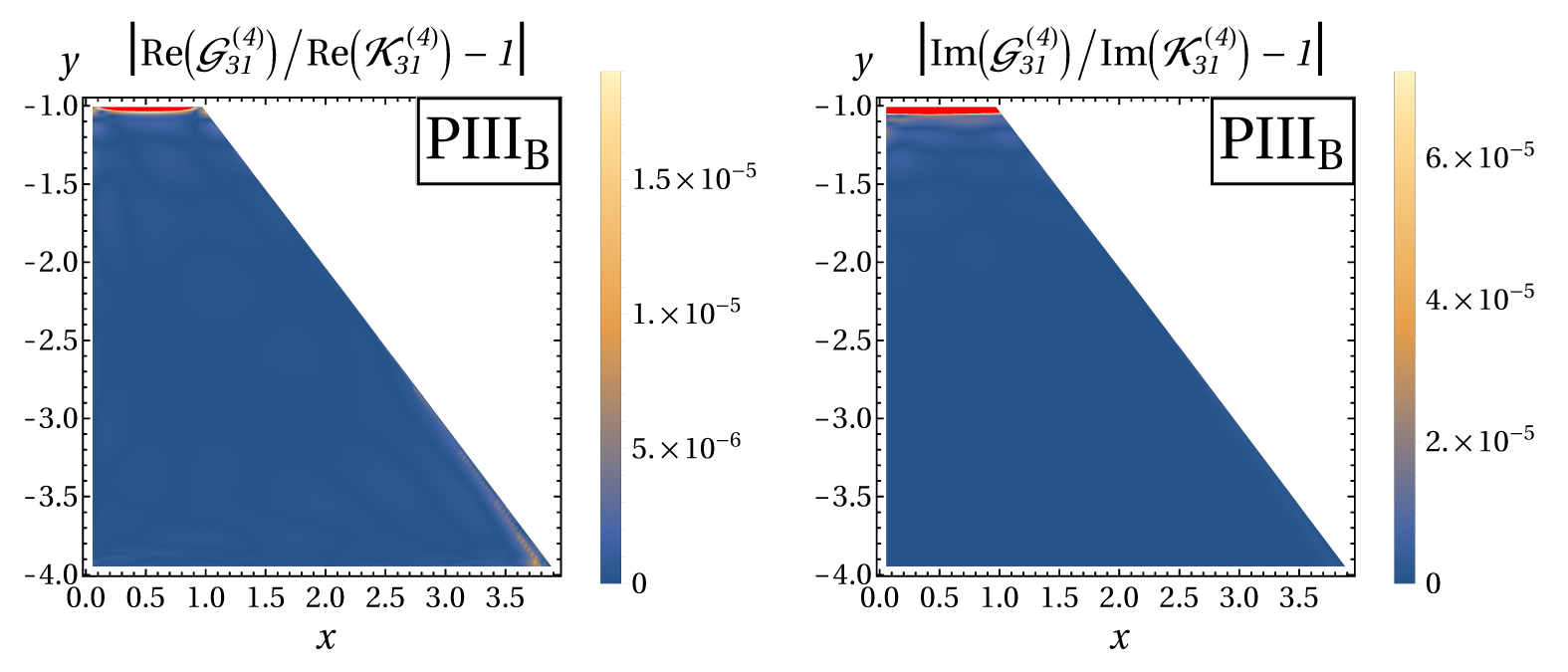}
		\label{PIII_err_B_31w4}
	\end{subfigure}\vspace{0.5cm}
	\caption{Heat map of the relative fitting error for the relevant weight of integral 31 of topology $\textnormal{P}_\textnormal{III}$ in region \textnormal{B}.}
	\label{PIII_err_B_31}
\end{figure}

\begin{figure}[H]
	\centering	
	\begin{subfigure}{1\linewidth}
		\centering
		\includegraphics[width=0.8\linewidth]{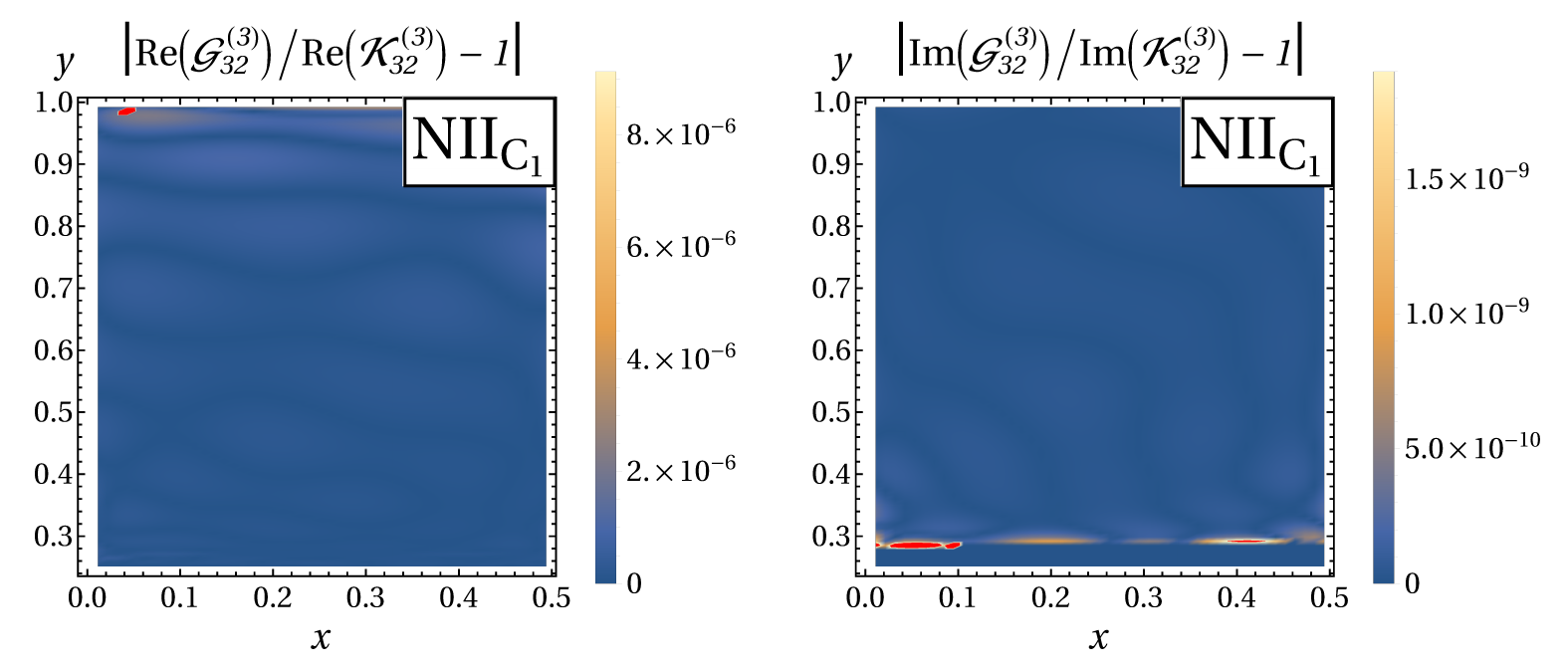}
		\label{NII_err_32w3}
	\end{subfigure}\vspace{0.2cm}	
	\begin{subfigure}{1\linewidth}
		\centering
		\includegraphics[width=0.8\linewidth]{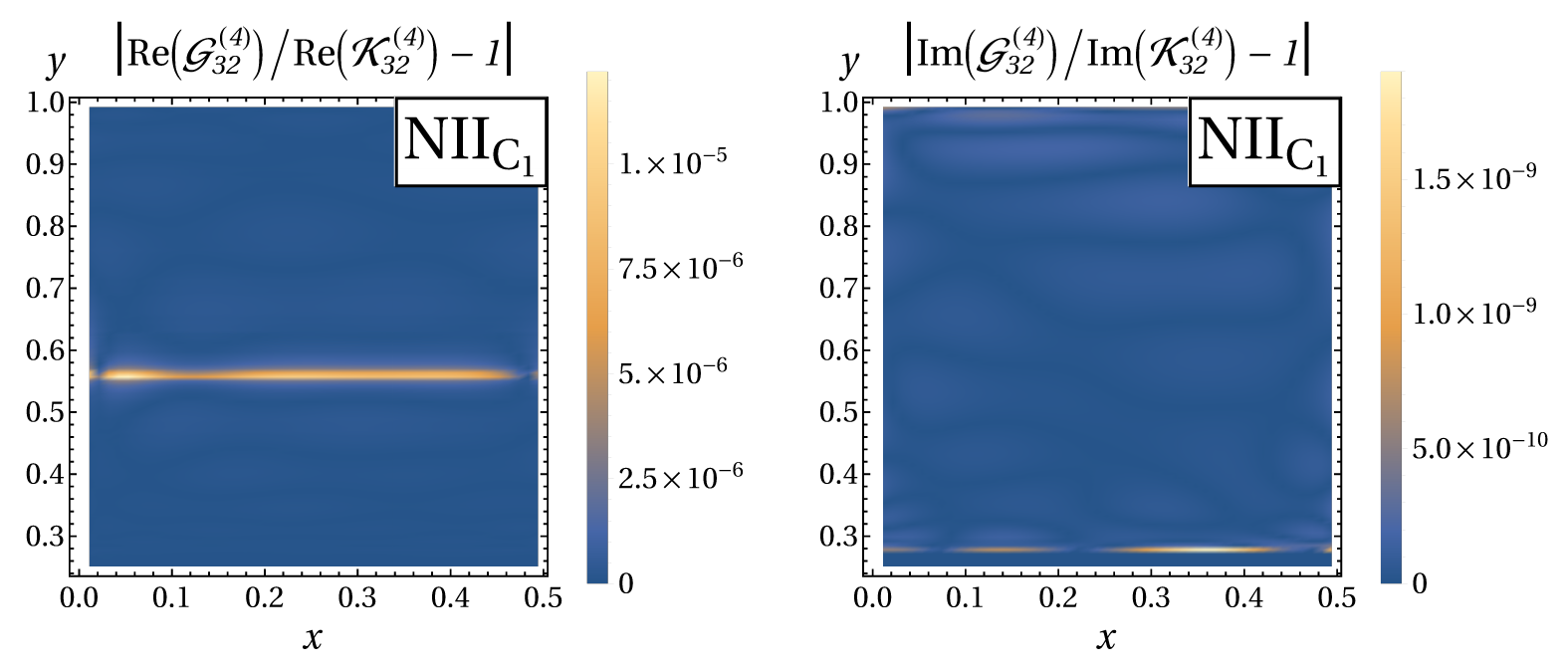}
		\label{NII_err_32w4}
	\end{subfigure}\vspace{0.2cm}	
	\caption{Heat map of the relative fitting error for the relevant weight of integral 32 of topology $\textnormal{N}_\textnormal{II}$ in region $\textnormal{C}_{\textnormal{1}}$.}		
	\label{NII_err_C1_32}
\end{figure}

\clearpage
\section{Results and Phenomenology}
\label{sec:Results}
\subsection{Overview}

In this section we use the calculation described and validated in the previous sections to perform a phenomenological study at the LHC. We have implemented the results into \texttt{MCFM} version 9.3~\cite{Campbell:2011bn,Campbell:2015qma} making particular usage of the existing architecture for diphoton production~\cite{Campbell:2011bn,Campbell:2016yrh}. \texttt{MCFM} provides the phase space generation, phase space cuts and PDF evaluation, for which we use the CT14NNLO PDF set~\cite{Dulat:2015mca}. For simplicity we focus on only the contributions arising from the $gg$ initiated channels, and define the $gg \rightarrow \gamma\gamma$ as our LO topology. We postpone a full phenomenological analysis (including the $q\overline{q}$ channels) to a later publication, once the full impact of the missing third generation, and other EW effects can be fully included. The reader might be interested to recall that the $gg$-initiated channel contributes around 25\% to 33\% of the $q\overline{q}$ (LO) result~\cite{Campbell:2011bn}.
Our results are determined using the following input parameters, 
\begin{itemize}
	\item $\sqrt{s} = 14\,\text{TeV}$,
	\item $M_W = 91.1876\,\text{GeV}$, 
	\item $M_Z = 80.3850\,\text{GeV}$, 
	\item $M_H = 125.0\,\text{GeV}$, 
	\item $\mu_F = \mu_R= M_Z\,\text{GeV}$, 
	\item $\alpha\left(0 \right) = 1/137.03599911$,
	\item $\alpha(M_Z) = 0.00775487$,
	\item CKM matrix $V_{ii} = 1$ for $i = u,d$, $V_{ij} = 0$ otherwise.
\end{itemize} 	
Additionally, we implemented the following kinematic cuts:
\begin{itemize}
	\item $p_T^{\gamma} > 30\,\text{GeV}$,
	\item $|\eta^\gamma| < 2.5$, 
	\item $R_{\gamma\gamma} = 0.4$
\end{itemize}
Our calculation naturally decomposes into two regions, a ``Higgs'' window $M_Z < m_{\gamma\gamma} < 2M_W$ and a ``high-energy'' window $m_{\gamma\gamma} > 2M_W$. In the first region the P$_\text{III}$ and  N$_\text{II}$ type topologies which contain two photons radiated from a $W$ box are below threshold. This region is also the region of interest for Higgs-interference studies. The high-energy regime occurs above the $2M_W$ threshold and, since the all contributions are above threshold one might expect a change in the phenomenology in this region. Motivated by these differences we study the NLO-EW corrections in the two different invariant mass windows  
separately. Since we do not yet fully include the third generation we cut the the invariant mass off at $500$ GeV. So that our two regions correspond, numerically, to $90\,\text{GeV} < m_{\gamma\gamma} < 2M_W$ and $2M_W < M_{\gamma\gamma} < 500\,\text{GeV}$.

\subsection{Cross sections}

In Tables~\ref{tab:cross-section_lt4}-~\ref{tab:cross-section_gt4}, we report the LO and NLO-EW cross sections for the gluon-induced diphoton process for the two invariant mass regions. The inclusion of the NLO electroweak corrections results in an enhancement of approximately $2.43\%$ and $1.62\%$ respectively.
\begin{table}[H]
	\centering
	\begin{tabular}{ccc}
		\hline
		\textbf{Order} & \textbf{Cross Section [pb]} & \textbf{Uncertainty [pb]} \\
		\hline
		$\tilde{\sigma}_{gg\to\gamma\gamma}^{(\text{LO})}$     & $2.176$ & $\pm\,0.002$ \\
		$\tilde{\sigma}_{gg\to\gamma\gamma}^{(\text{NLO-EW})}$    & $2.229$ & $\pm\,0.003$ \\
		\hline
	\end{tabular}
	\caption{Comparison of LO and NLO-EW renormalized cross-section results from \texttt{MCFM} in the region $ 90 \le  m_{\gamma\gamma} \le 2M_W$.}
	\label{tab:cross-section_lt4}
\end{table}

\begin{table}[H]
	\centering
	\begin{tabular}{ccc}
		\hline
		\textbf{Order} & \textbf{Cross Section [pb]} & \textbf{Uncertainty [pb]} \\
		\hline
		$\tilde{\sigma}_{gg\to\gamma\gamma}^{(\text{LO})}$     & $0.39119$ & $\pm\,0.00001$ \\
		$\tilde{\sigma}_{gg\to\gamma\gamma}^{(\text{NLO-EW})}$    & $0.39751$ & $\pm\,0.00006$ \\
		\hline
	\end{tabular}
	\caption{Comparison of LO and NLO-EW renormalized cross-section results from \texttt{MCFM} in the region $2M_W < m_{\gamma\gamma} < 500\,\text{GeV}$.}
	\label{tab:cross-section_gt4}
\end{table}

\subsection{Differential Distributions }

Next we turn our attention to differential distributions, again focusing on the two invariant mass-regions separately. Figure~\ref{fig:M_gaga} displays the invariant mass distributions for the two regions. In order to gleam further insight into the nature of the corrections we breakdown the EW correction into component pieces (defined in Eq.~\ref{LO2},~\ref{QED2} and \ref{Z2} respectively), which are presented in the middle panel. 
Although such a breakdown is not physical (since terms can be shuffled between the pure two-loop correction and the counter-term by choice of renormalization scheme), such a decomposition is interesting, viewed through the lens of a specific UV setup, to understand the importance of the various contributions in different phase space regions. In particular quickly glancing at the figure illustrates the hierarchy between the various contributions: the $W$-mediated diagrams (including also the $\phi$ Goldstones) dominate over the other contributions, with the counter-terms (CT) and the QED corrections contributing approximately the same, and the neutral current $Z$ diagrams being more than two order of magnitudes smaller and negative in size.  The lower panel in the figures shows the differential $K$-factor in going from LO to NLO-EW, in the lower-energy regions we observe that the EW correction is essentially a flat $\sim2.5\%$ correction to the LO contribution. This changes above the $2M_W$ threshold, here the contributions begin to diminish and eventually $(\sim 300$ GeV) begin to decrease the LO result. However, care should be taken when interpreting the results in this region of phase space since here the impact of the top-quark is expected to be significant. The investigation of the very high-energy tail $>500$ GeV is therefore motivated, however we postpone this until completion of the third-generation contributions. 

\begin{figure}[H]
	\centering
	\includegraphics[width=0.485\linewidth]{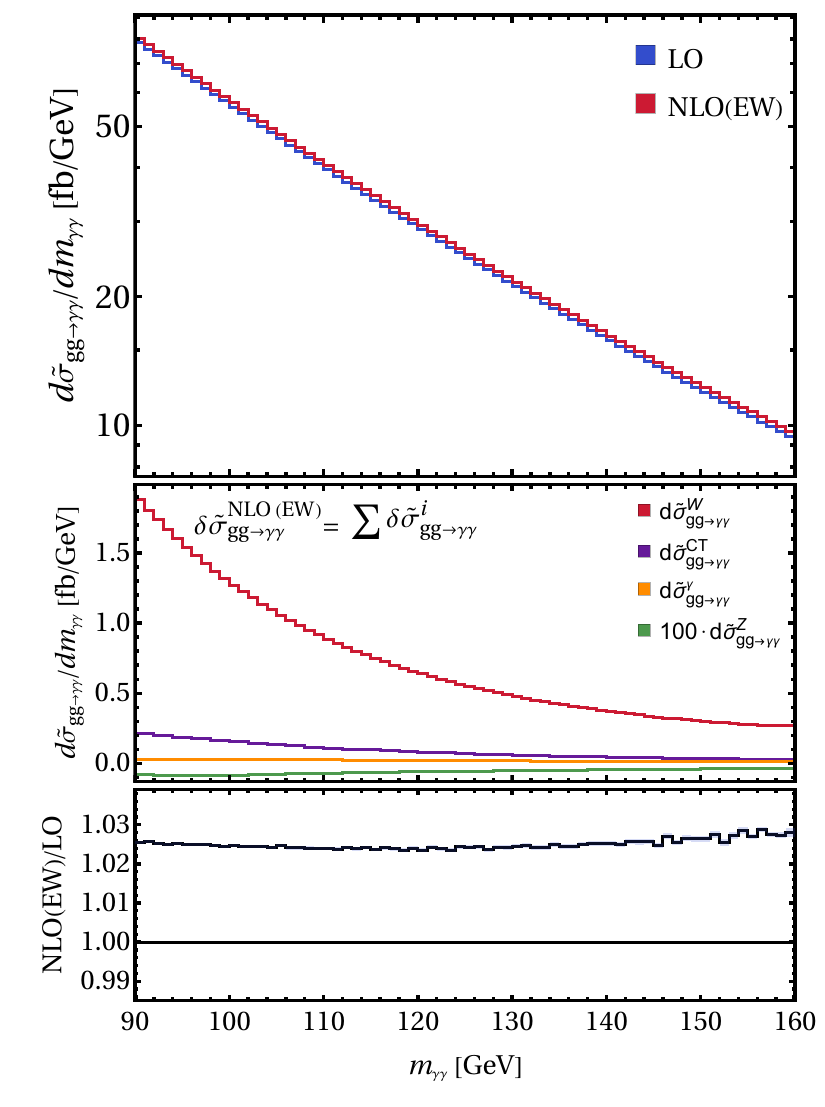}
	\includegraphics[width=0.485\linewidth]{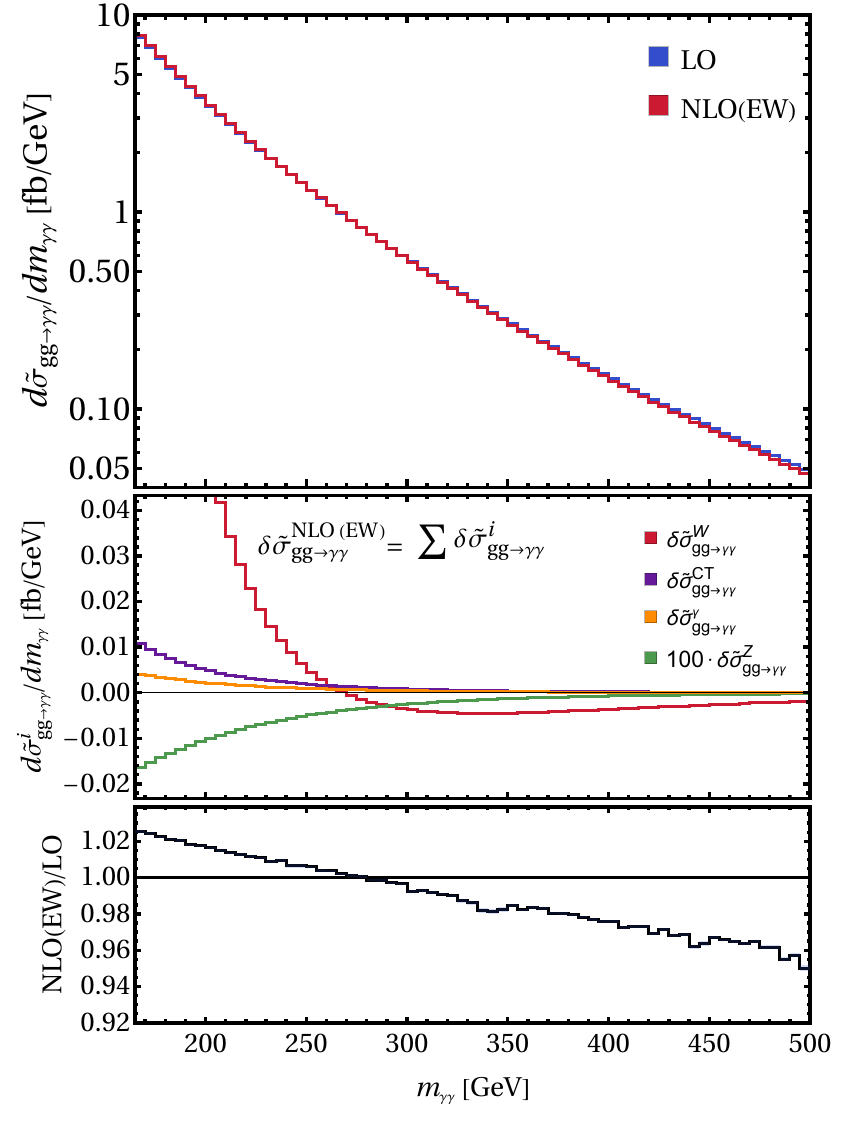}
	\caption{Differential distributions for the invariant mass of the photon pair in the kinematic region $m_{\gamma\gamma} < 2M_W$ (left) and $m_{\gamma\gamma} > 2 M_W$ (right). The middle panel shows the contributions arising from the different mediating bosons that enter the calculation.
		Lower panel present the ratio between the NLO-EW and the LO distributions with the propagation of Monte Carlo error highlighted in shaded blue.}
	\label{fig:M_gaga}
\end{figure}

Figure~\ref{fig:pt_ga} shows the differential cross section with respect to the transverse momentum of the hardest photon, $d\tilde{\sigma}/dp_T^\gamma$. 
Again we present the low-energy and high-energy regions separately, with the lower-energy distribution on the left. 
Given the absence of real radiation, the final state consists of exactly two photons with no additional partons. As a result, the photons are produced strictly back-to-back in the transverse plane, and their transverse momenta are equal in magnitude: $p_T^{\gamma_1} = p_T^{\gamma_2}$. Consequently, the diphoton system has zero transverse momentum, and the $p_T^\gamma$ spectrum corresponds to the common transverse momentum of either photon.

The LO distribution exhibits a distinct peak around $p_T^\gamma \sim 45(85) ~\text{GeV}$ for the low-(high-)energy region. This peak arises from the combined effects of phase-space constraints and the imposed cuts on the diphoton invariant mass. Given the invariant mass cut $90~\text{GeV} \leq m_{\gamma\gamma} \leq 160~\text{GeV}$, symmetric photon configurations near the lower edge of the invariant mass window dominate. Using the approximate kinematic relation for back-to-back photons,
\begin{equation}
	m_{\gamma\gamma} \approx 2 p_T^\gamma \cosh\left(\frac{\Delta y}{2}\right),
\end{equation}
and assuming a typical rapidity separation $|\Delta y| \sim 1$, the corresponding photon transverse momentum peaks at the minimum $m_{\gamma\gamma}/2$. The EW corrections induce a flat correction of around 2\% in the Higgs region, with a slightly more structure in the high energy region, although the impact here also remains close to $2\% $ .

\begin{figure}[H]
	\centering
	\includegraphics[width=0.48\linewidth]{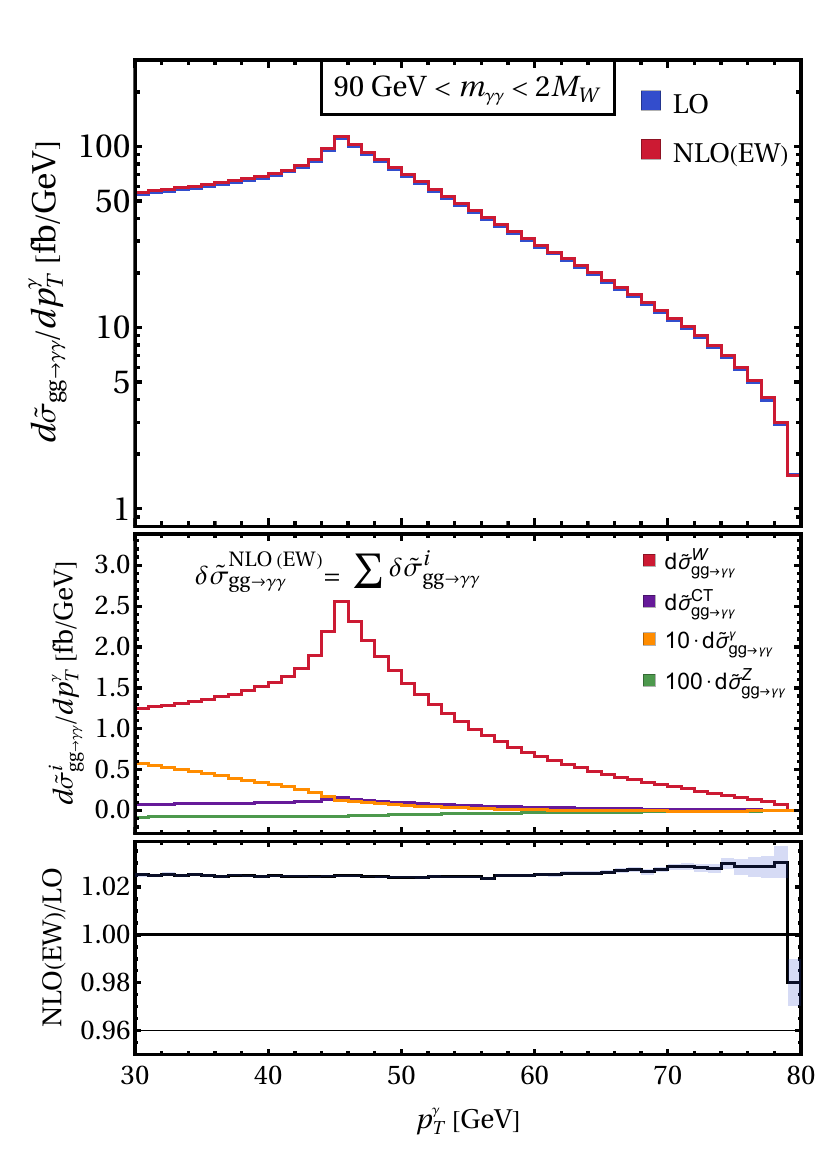}
	\includegraphics[width=0.48\linewidth]{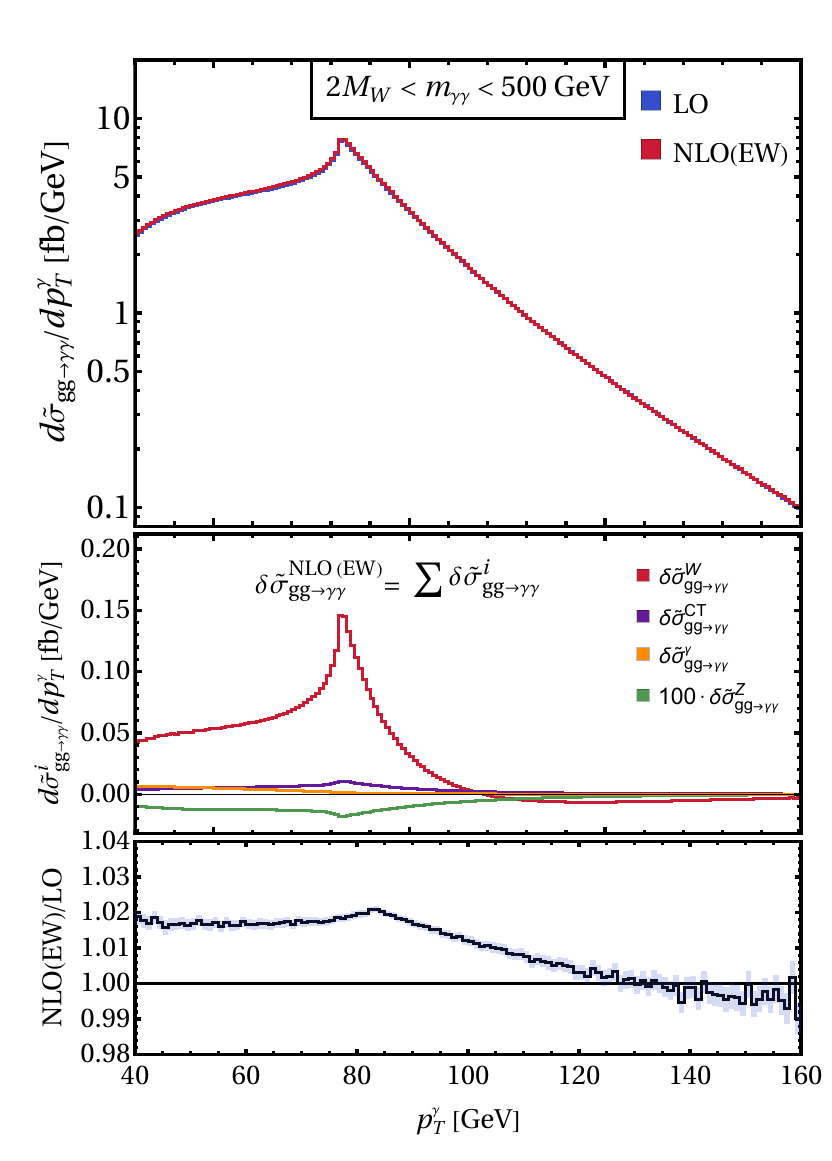}
	\caption{Differential distributions for the transverse momentum of the hardest photon in the kinematic region $m_{\gamma\gamma} < 2M_W$ (left) and $m_{\gamma\gamma} > 2 M_W$ (right). The middle panel shows the contributions arising from the different mediating bosons that enter the calculation.\
		Lower panel present the ratio between the NLO-EW and the LO distributions with the propagation of Monte Carlo error highlighted in shaded blue.}
	\label{fig:pt_ga}
\end{figure}
\begin{figure}[H]
	\centering
	\includegraphics[width=0.48\linewidth]{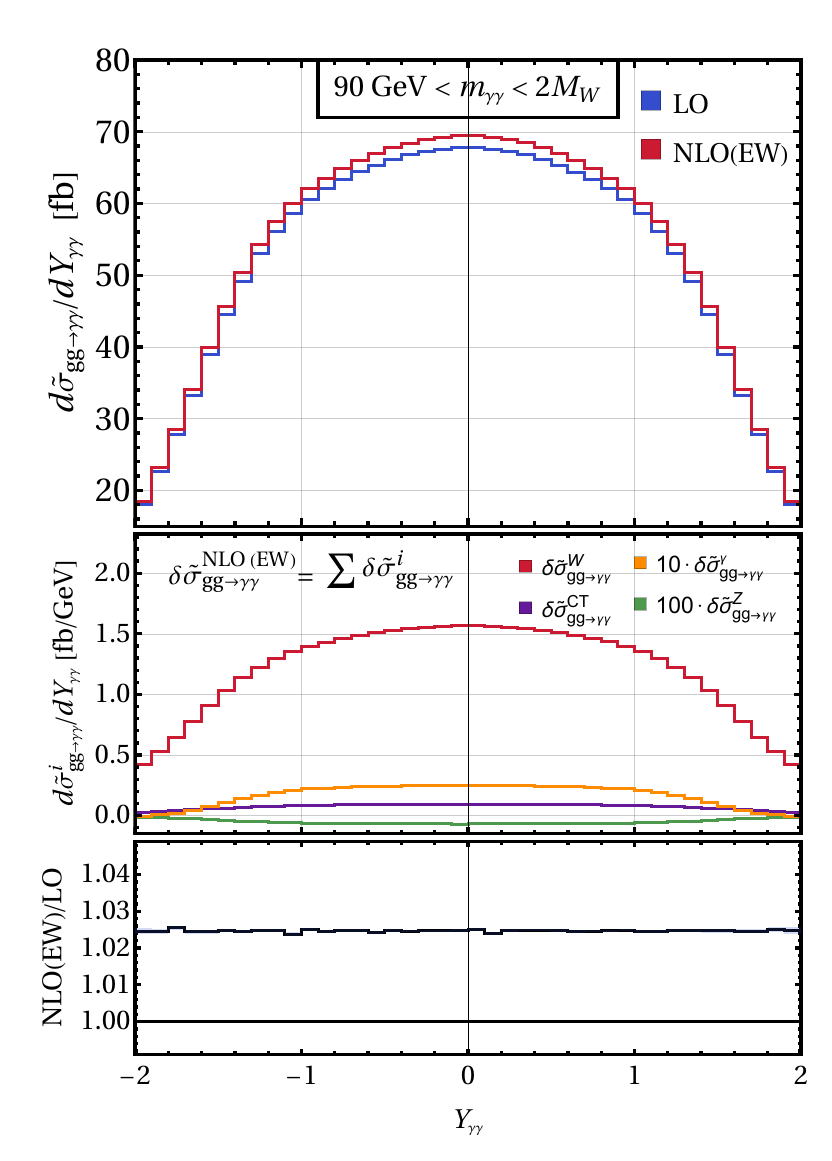}
	\includegraphics[width=0.48\linewidth]{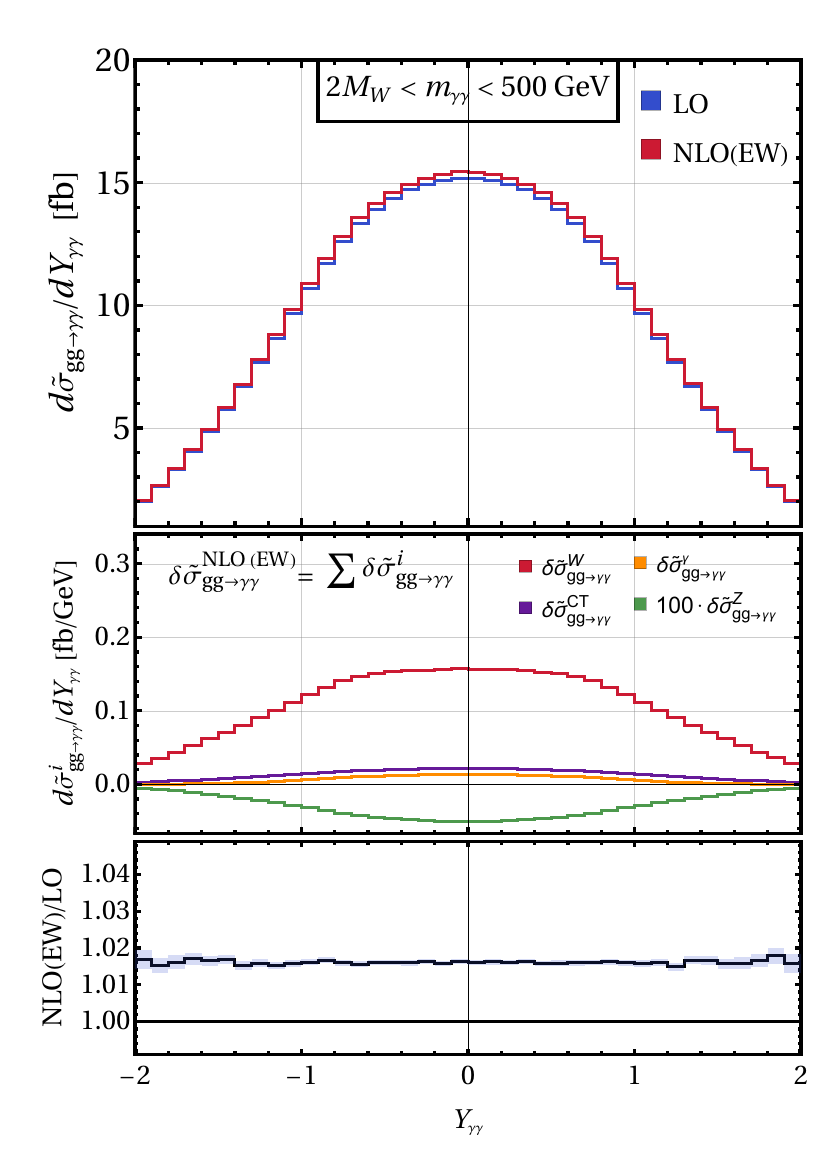}
	\caption{Rapidity distributions for the $\gamma\gamma$ pair in the kinematic region $2M_W < m_{\gamma\gamma} < 500\,\text{GeV}$. The central panel shows the contribution between the different mediator diagrams that enter the calculation.  Below, the ratio between the NLO and the LO distributions with the propagation of Monte Carlo error highlighted in shaded blue.}
	\label{fig:Y_gaga}
\end{figure}

Finally, in Figure~\ref{fig:Y_gaga} we present the rapidity distribution for the photon pair $Y_{\gamma\gamma}$, again keeping the two regions separate. Since shape of the rapidity distribution is mostly sensitive to the partonic PDFs, which are the same for us at LO and NLO-EW there is very little change in shape at NLO-EW and the corrections come in as a flat 2\% correction across the phase space. 


\section{Conclusions}

In this paper we have presented the calculation of the electroweak corrections to $gg\rightarrow \gamma\gamma$, focusing on the impact of the first two (massless) generations of quarks. Production of diphotons through gluon fusion nominally is a NNLO QCD correction to the $q\overline{q}$ initiated channel, however the large flux of gluons in the colliding process enhances this contribution such that it contributes a significant ($\sim 1/3)$ amount of the total cross section.  The production of photon pairs is a process of great importance at the LHC primarily given its role in the continued studies of the Higgs boson. One fascinating ongoing study at the LHC is the interference of the photons produced through the decay of a Higgs boson with the continuum $gg$ production. Given the large data set of the HL-LHC a precision measurement of the cross section in the vicinity of the Higgs may yield constraints on the Higgs width. This interference has been shown to be sensitive to higher order corrections in QCD, studying the impact of the EW corrections to the interference is a hence a motivating principle of this work. 

In order to calculate the electroweak corrections we used tensor projectors to define a collection of eight form-factors. These form-factors contain thousands of scalar integrals which we reduced using well-known Integration by parts identities to a much smaller collection of Master Integrals (MIs). We completed the calculation of these MIs in-terms of Chen iterated integrals in a previous campion paper. In order to evaluate the MIs efficiently in a Monte Carlo code over a wide phase space we used our analytic results to produce a grid. Then we employed fitting functions to approximate the expansion of the MIs as a series in the dimensional regulating parameter $\epsilon$. We investigated the validity of our fits across the phase space, finding agreement at around $10^{-6}$ across the bulk of the phase space. Additionally we computed the contribution from one topology analytically and using our fitting procedure, finding excellent agreement. 

We implemented our results into \texttt{MCFM} and studied the impact of the correction on the LO $gg\rightarrow \gamma\gamma$ process, finding a correction of around $2.43\%$ in the ``low''-invariant mass region  $(< 2M_W$), and around $1.62\%$ in the ``high''-invariant ($ > 2M_W$) regions, with some differing phenomenology at higher invariant masses. 
After the completion of the third generation it will be extremely interesting to study the effects of the electroweak corrections on the interference of the Higgs-continuum $gg$ production. We leave this study to a future publication. 

\section*{Acknowledgments}

The authors thank Doreen Wackeroth and Daniele Gaggero for useful discussions. 
This work was supported by the National Science Foundation through
award NSF-PHY-2310363, GF also acknowledges support from the Swiss National Science Foundation (SNF) under contract 200021-231259. Support provided by the Center for Computational Research at the University at Buffalo. 

\bibliographystyle{JHEP}

\bibliography{gg2gaga}

\end{document}